\documentclass[prd,aps,nofootinbib,floats,letterpaper,floatfix,superscriptaddress,groupedaddress,eqsecnum,twocolumn]{revtex4}
\usepackage[usenames]{color}
\usepackage{epsfig}
\usepackage{amssymb,amsmath}

\begin{document}
\def\be{\begin{equation}}
\def\ee{\end{equation}}
\def\beq{\begin{eqnarray}}
\def\eeq{\end{eqnarray}}
\def\LL{\left[\left[}
\def\RR{\right]\right]}

\def\msun{M_\odot}

\def\nn{\nonumber}
\def\lappreq{\! \stackrel{\scriptscriptstyle <}{\scriptscriptstyle
\sim}\!}
\def\gappreq{\!\stackrel{\scriptscriptstyle >}{\scriptscriptstyle \sim}\!}
\def\ver{\vskip 12pt}

\title{Gravitational wave signatures of the absence of an event horizon. \\ 
I. Nonradial oscillations of a thin-shell gravastar} 

\author{Paolo Pani} \email{paolo.pani@ca.infn.it} \affiliation{Dipartimento di
  Fisica, Universit\`a di Cagliari, and INFN sezione di Cagliari, Cittadella
  Universitaria 09042 Monserrato, Italy \\Currently at Centro Multidisciplinar
  de Astrof\'{\i}sica - CENTRA, Dept. de F\'{\i}sica, Instituto Superior
  T\'ecnico, Av. Rovisco Pais 1, 1049-001 Lisboa, Portugal}

\author{Emanuele Berti} \email{berti@phy.olemiss.edu} \affiliation{Department
  of Physics and Astronomy, The University of Mississippi, University, MS
  38677-1848, USA} \affiliation{Theoretical Astrophysics 350-17, California
  Institute of Technology, Pasadena, CA 91125, USA}

\author{Vitor Cardoso} \email{vitor.cardoso@ist.utl.pt}
\affiliation{Department of Physics and Astronomy, The University of
  Mississippi, University, MS 38677-1848, USA} \affiliation{Centro
  Multidisciplinar de Astrof\'{\i}sica - CENTRA, Dept. de F\'{\i}sica,
  Instituto Superior T\'ecnico, Av. Rovisco Pais 1, 1049-001 Lisboa, Portugal}

\author{Yanbei Chen} \email{yanbei@tapir.caltech.edu} 
\affiliation{Theoretical Astrophysics 350-17, California Institute of
  Technology, Pasadena, CA 91125, USA}

\author{Richard Norte} \email{norte@caltech.edu} 
\affiliation{Theoretical Astrophysics 350-17, California Institute of
  Technology, Pasadena, CA 91125, USA}

\begin{abstract}
  Gravitational waves from compact objects provide information about their
  structure, probing deep into strong-gravity regions.  Here we illustrate how
  the presence or absence of an event horizon can produce qualitative
  differences in the gravitational waves emitted by ultra-compact objects. In
  order to set up a straw-man ultra-compact object with no event horizon, but
  which is otherwise almost identical to a black hole, we consider a
  nonrotating thin-shell model inspired by Mazur and Mottola's gravastar,
  which has a Schwarzschild exterior, a de Sitter interior and an infinitely
  thin shell with finite tension separating the two regions.  As viewed from
  the external space-time, the shell can be located arbitrarily close to the
  Schwarzschild radius, so a gravastar might seem indistinguishable from a
  black hole when tests are only performed on its external metric.  We study
  the linearized dynamics of the system, and in particular the junction
  conditions connecting internal and external gravitational perturbations. As
  a first application of the formalism we compute polar and axial oscillation
  modes of a thin-shell gravastar.  We show that the quasinormal mode spectrum
  is completely different from that of a black hole, even in the limit when
  the surface redshift becomes infinite.  Polar QNMs depend on the equation of
  state of matter on the shell and can be used to distinguish between
  different gravastar models. Our calculations suggest that low-compactness
  gravastars could be unstable when the sound speed on the shell $v_s/c\gtrsim
  0.92$.
\end{abstract}

\maketitle

\section{Introduction}

Black holes (BHs), once considered an exotic mathematical solution to
Einstein's field equations, have now been widely accepted as astronomical
objects
\cite{Melia:2007vt,Narayan:2005ie,Mueller:2007rz,Berti:2009kk}. Stellar-mass
BHs are believed to be the final stage of the evolution of sufficiently
massive stars. Massive BHs seem to populate the center of many galaxies at low
redshift, and must have played an important role in the formation of structure
in the universe. Most evidence supporting the astrophysical reality of BHs
comes from the weak-gravity region, i.e. from observations probing the
space-time several Schwarzschild radii away from the event horizon. Attempts
to rule out possible alternatives to BHs usually rely on general relativity
being the correct theory of gravity, and/or on constraints on the equation of
state of matter at high densities.  For massive BHs the most precise
measurements so far come from observations of stellar proper motion at the
center of our own galaxy, indicating the presence of a ``dark object'' of mass
$M\simeq (4.1\pm 0.6)\times 10^6M_\odot$ \cite{Ghez:2008ms}. Recent millimeter
and infrared observations of Sagittarius A$^*$, the compact source of radio,
infrared and X-ray emission at the center of the Milky Way, infer an intrinsic
diameter of 37$^{+16}_{-10}$ microarcseconds, even {\em smaller} than the
expected apparent size of the event horizon of the presumed BH
\cite{Doeleman:2008qh}.  Some of the exotic alternatives to a BH (such as
``fermion balls'') are incompatible with the observations
\cite{Schodel:2002py} and any distribution of individual objects within such a
small region (with the possible exception of dark matter particles or
asteroids, which however should be ejected by three-body interactions with
stars) would be gravitationally unstable
\cite{Maoz:1998,ColemanMiller:2005tg}.  In a recent attempt to probe the event
horizon, Broderick and Narayan have analyzed the observations of
Ref.~\cite{Doeleman:2008qh}. If the object at the center of our galaxy had a
surface it would be hot enough to glow with a steady emission of infrared
light, but no such glow has been detected \cite{Broderick:2009ph}. This and
similar arguments are inevitably dependent on the gas distribution and on
details of the accretion process, and they really set lower limits on the
gravitational redshift corresponding to the hypothetical surface replacing the
event horizon (see e.g. \cite{Psaltis:2008bb} for a review).  Indeed, some
hold the view that an observational proof of the existence of event horizons
based on electromagnetic observations is fundamentally impossible
\cite{Abramowicz:2002vt}.  For these reasons, model-independent tests of the
strong-field dynamics of BHs and studies of the gravitational radiation
signatures of event horizons are necessary to confirm or disprove the BH
paradigm \cite{Kundt:2009vz,Visser:2009xp}.

Gravitational wave detectors offer a new way of observing BHs, complementing
the wealth of information from present electromagnetic observations
\cite{Berti:2009kk}.  As first proposed by Ryan
\cite{Ryan:1995wh,Ryan:1997hg}, an exquisite map of the external space-time of
BHs (outside the innermost stable orbit, if there is any) can be constructed
by observing the gravitational waveform emitted when a small compact object
spirals into the putative supermassive BH at the center of a galaxy with the
Laser Interferometer Space Antenna (LISA).  As an extension of Ryan's work, Li
and Lovelace considered the small object's tidal coupling with the central
object and showed that by studying details of radiation reaction, information
about the space-time region {\it within} the orbit can also be obtained
\cite{Li:2007qu}.
Ryan's proposal to map space-times using inspiral waveforms is promising, but
the data analysis task is affected by a ``confusion problem'': the possibility
of misinterpreting truly nonKerr waveforms by Kerr waveforms with different
orbital parameters \cite{Collins:2004ex,Glampedakis:2005cf}. This ambiguity
was shown to be resolvable if the orbit is known to be circular
\cite{Ryan:1995wh} or if one only probes the mass, spin and quadrupole moment
of the object using waveforms generated in the weak-gravity region
\cite{Barack:2007}.  A different approach to test the BH nature of
ultra-compact objects is based on measuring several of their free oscillation
frequencies (``ringdown waves'') and comparing them with the quasinormal mode
(QNM) spectrum of BHs \cite{Dreyer:2003bv}. These tests of the ``no-hair
theorem'' require a signal-to-noise ratio which could be achieved by advanced
Earth-based gravitational wave interferometers and they are one of the most
promising science goals of LISA
\cite{Berti:2005ys,Berti:2007zu,Sathyaprakash:2009xs}.

In this paper we explore how to test possible alternatives to the BH
paradigm. We retain the ``conservative'' assumption that general relativity is
the correct theory of gravity and we focus on possible tests of the existence
(or absence) of an event horizon.

Theorists conceived several families of compact objects with no event
horizons. For example, boson stars are horizonless compact objects based on
plausible models of particle physics at high densities, and they are (still)
compatible with astrophysical observations
\cite{Schunck:2003kk,Torres:2000dw}.  Being indistinguishable from BHs in the
Newtonian regime, boson stars are good ``strawmen'' for supermassive BHs. The
space-time of nonrotating spherical boson stars can approximate arbitrarily
well a Schwarzschild geometry even close to the event horizon, and being very
compact it is not easily distinguishable from a BH by electromagnetic
observations \cite{Torres:2000dw,Guzman:2005bs}.  Building on Ryan's proposal,
Kesden {\it et al.} showed that the inspiral of a small compact object into a
nonrotating boson star will emit a rather different gravitational waveform at
the end of the evolution, when the small object falls into the central
potential well of the boson star instead of disappearing into the event
horizon of a BH \cite{Kesden:2004qx}. Several authors have computed the QNM
spectrum of boson stars, showing that it is remarkably different from the QNM
spectrum of BHs and lending support to the feasibility of no-hair tests using
QNM measurements \cite{Yoshida:1994xi,Balakrishna:2006ru,Berti:2006qt}.

Another proposed alternative to massive BHs, which we shall focus on in this
paper, are the so-called {\it gravastars} \cite{Mazur:2001fv}. The gravastar
model assumes that the space-time undergoes a quantum vacuum phase transition
in the vicinity of the BH horizon.  The model effectively replaces the BH
event horizon by a transition layer (or shell) and the BH interior by a
segment of de Sitter space \cite{Chapline:2000en,Mazur:2004fk}. Mazur and
Mottola argued for the thermodynamic stability of the model.  A dynamical
stability analysis by Visser and Wiltshire \cite{Visser:2003ge} confirmed that
a simplified version of the gravastar model by Mazur and Mottola is also
stable under radial perturbations for {\it some} physically reasonable
equations of state for the transition layer. Chirenti and Rezzolla
\cite{Chirenti:2007mk} first considered nonradial perturbations of gravastars,
restricting attention to the relatively simple case of oscillations with axial
parity. They computed the dominant axial oscillation modes and found no
instabilities. In analogy with previous studies of the oscillation modes of
boson stars \cite{Yoshida:1994xi,Balakrishna:2006ru,Berti:2006qt}, they
confirmed that the axial QNM spectrum of gravastars can be used to discern a
gravastar from a BH. In the thin-shell limit, the axial QNM frequencies of
Ref.~\cite{Chirenti:2007mk} and our own calculations recover Fiziev's
calculation of the axial QNMs of ultracompact objects with a totally
reflecting surface \cite{Fiziev:2005ki}.

In this paper we study the stability with respect to nonradial oscillations
of a simplified ``thin-shell'' gravastar, retaining most of the essential
features of the original model. We consider both axial perturbations
(reproducing and extending the results of Ref.~\cite{Chirenti:2007mk}) and
polar perturbations. In the polar case, the matching of interior and exterior
perturbations at the gravastar shell requires a more careful analysis because
(unlike the axial case) polar perturbations of spherical objects actually
induce motions of matter, which in turn couples back to gravitational
perturbations. For this reason, nonspherical polar perturbations provide a
more stringent test on the gravastar's overall stability.  Polar perturbations
are also crucial in studying the dynamics of objects orbiting the hypothetical
gravastar. 
In order to treat polar perturbations of a nonrotating thin-shell gravastar we
set up a rather generic formalism combining standard perturbation theory (in
the Regge-Wheeler gauge) with Israel's junction conditions
\cite{Israel:1966rt}.  The formalism can be applied to gravitational
perturbations of any spherically symmetric space-time characterized by regions
with different cosmological constants separated by infinitely thin shells with
finite surface energy and tension. Because matter is concentrated on these
shells, the junction conditions deduced here will be sufficient in describing
the linear dynamics of matter.  Quite predictably, these conditions depend on
the equation of state of the shell.  As an application of the formalism we
study the QNM spectrum of polar and axial perturbations of gravastars,
exploring the nonradial stability of these objects. Polar QNMs (unlike axial
QNMs) depend on the equation of state of matter on the shell: they can be used
not only to distinguish a gravastar from a BH, but also to distinguish between
different gravastar models. We also find that the imaginary part of some QNMs
seems to have a zero crossing when the gravastar is not very compact and the
speed of sound on the shell is superluminal, suggesting that some gravastar
models may be unstable under nonradial perturbations.  In a companion paper we
will apply the formalism developed in this paper to study gravitational
radiation from compact objects inspiraling into nonrotating gravastars.

The plan of the paper is as follows. In Section \ref{sec:equilibrium} we
review our static thin-shell gravastar model. Section \ref{sec:pert} sketches
the calculation of axial and polar gravitational perturbations and of the
matching conditions at the gravastar shell. Details of the matching procedure
are provided in Appendix \ref{app:match}, and details of the QNM calculation
are given in Appendix \ref{app:cf}. Our numerical results for the polar and
axial QNM spectra are presented in Section \ref{sec:results} and supported by
analytical calculations in the high-compactness limit in Appendix
\ref{app:highC}.

We use geometrical units ($G=c=1$).  The Fourier transform of the perturbation
variables is performed by assuming a time dependence of the form $e^{i\omega
  t}$.  Greek indices $(\mu\,,\nu\,,\dots)$ refer to the four-dimensional
space-time metric. Latin indices $(i\,,j\,,\dots)$ refer to the
three-dimensional space-time metric on the shell. Latin indices at the
beginning of the alphabet $(a\,,b\,,\dots)$ refer to the spatial metric on a
two-sphere.

%%%%%%%%%%%%%%%%%%%%%%%%%%%%%%%%%%%%%%%%%%%%%%%%%%%%%%%%%%%%%%%%%%%%%%%%%%%%%%%%%%%%%%%%%%%%%%%%%%%%%%%%%%
\section{Equilibrium model}\label{sec:equilibrium}
%%%%%%%%%%%%%%%%%%%%%%%%%%%%%%%%%%%%%%%%%%%%%%%%%%%%%%%%%%%%%%%%%%%%%%%%%%%%%%%%%%%%%%%%%%%%%%%%%%%%%%%%%%

The metric for a static thin-shell gravastar has the form
\cite{Visser:2003ge,Chirenti:2007mk}
\begin{equation} 
ds_0^2 = - f(r) dt^2+\frac{1}{h(r)}dr^2 + r^2(d\theta^2+\sin^2\theta d\varphi^2)\,.\label{eq:g0}
\end{equation}
Here
\begin{equation}
f(r) = \left\{
\begin{array}{ll}
\displaystyle
h(r)= 
1-\frac{2M}{r} \,, & r>a \,,\\
\\
\displaystyle 
\alpha\,h(r)=
\alpha
\left(
1-\frac{8\pi\rho}{3}r^2
\right)
\,, & r<a \,,
\end{array}
\right.
\label{fr}
\end{equation}
where $M$ is the gravastar mass measured by an outside observer, and
$\rho=3M/(4\pi a^3)$ is the ``energy density'' of the interior region.  The
coordinate system $(t,r,\theta,\varphi)$ has been chosen in such a way that
the thin shell occupies a coordinate sphere with $r=a$.  The space-time
reduces to de Sitter for $r<a$, and to Schwarzschild for $r>a$. The junction
conditions on the $r=a$ surface have already partially been chosen by
requiring the {\it induced metric} to be continuous across the shell, which
also dictates that $f(r)$ be continuous at $r=a$, or
\begin{equation}
\label{fcont}
1-\frac{2M}{a} = \alpha\left(1-\frac{8\pi\rho a^2}{3}\right)\,.
\end{equation}
In this paper we shall usually drop the dependence of $f(r)$ and $h(r)$ on
$r$.  From the jump in the radial derivatives of $f$ we could easily obtain
the two defining properties of this gravastar model: the surface energy
density $\Sigma$ and surface tension $\Theta$.
The junction conditions read \cite{Israel:1966rt}:
\begin{equation}
[[K_{ij}]] = 8\pi \left[\left[ S_{ij} -\gamma_{ij} \frac{S}{2}\right]\right]\,,
\end{equation}
where the symbol ``[[ ...]]'' gives the ``jump'' in a given quantity across
the spherical shell (or $r=a$), i.e.
\begin{equation}
[[A]] \equiv {A(a_+)-A(a_-)} \,.\label{eq:defjump}
\end{equation}
The indices $i$ and $j$ correspond to coordinates $t$, $\theta$, and $\varphi$
which parameterize curves tangential to the spherical shell, $K_{ij}=-\nabla_i
n_j$ is the extrinsic curvature, $n_\alpha = (0,1,0,0)/\sqrt{g^{rr}}$ is the
unit normal vector, and $S_{ij}$ is the surface stress-energy tensor
\begin{equation}
S_{ij} = (\Sigma -\Theta) u_i u_j  -\Theta \gamma_{ij}\,,
\end{equation}
where $u^\alpha = \sqrt{-1/g_{tt} } (1,0,0,0)$ (or $\vec{u} =
\sqrt{-1/g_{tt}}\, \vec{\partial}_t$) is the four-velocity of mass elements on
the shell and $\gamma_{\alpha\beta} = g_{\alpha\beta} - n_{\alpha}n_{\beta}$
is the induced 3-metric on the shell.  We then have
\begin{equation}
S_{ij} - \gamma_{ij}\frac{S}{2} = 
(\Sigma-\Theta) u_i u_j +\frac{\Sigma}{2}  \gamma_{ij}\,.
\end{equation}
In the static, spherically symmetric case,
\begin{equation}
\LL K_{ij} \RR =\left[ \frac{\sqrt{g^{rr}} }{2} g_{ij,r}\right]\,.
\end{equation}
Discontinuities in the metric coefficients are then related to the surface
energy and surface tension as \cite{Visser:2003ge}
\begin{equation}
[[\sqrt{h}]] =-4\pi a \Sigma\,,\quad 
\left[\left[ \frac{f'\sqrt{h}}{f}\right]\right]= 8\pi (\Sigma-2\Theta)\,. \label{eq:SigmaTheta}
\end{equation}
In order to summarize the above relations and reveal the independent
parameter space of a thin-shell gravastar we define
\begin{equation}
M_v \equiv \frac{4\pi \rho a^3}{3} \,, \quad
M_s \equiv 4\pi a^2 \Sigma\,,
\end{equation}
which would be the volume- and surface-energy contents of the gravastar. In
terms of $M_v$, $M_s$ and $a$, we can obviously solve for $\rho$, $\Sigma$,
and in addition we have
\begin{eqnarray}
M&=&M_v + M_s\sqrt{1-\frac{2M_v}{a}}+\frac{M_s^2}{2a}\,, \\
\alpha&=&\frac{1-2M/a}{1-2M_v/a}\,, \\
\Theta&=&\frac{1}{8\pi a}\left[
\frac{1-4M_v/a}{\sqrt{1-2M_v/a}}
-\frac{1-M/a}{\sqrt{1-2M/a}}
\right]\,.
\end{eqnarray}
As a consequence, gravastar types can be specified by the dimensionless
parameters $M_v/a$ and $M_s/a$.  In this paper we only consider a simplified
version of the original Mazur-Mottola gravastar, which has vanishing surface
energy ($\Sigma=0$) and
\begin{equation}
M=M_v,\quad 8\pi a \Theta=-\frac{3M/a}{\sqrt{1-2M/a}}\,.
\end{equation}
However it is convenient to keep our notation general enough, because
nonradial oscillations of a gravastar will in general produce nonzero
variations of the surface energy (i.e., $\delta \Sigma \neq 0$).

\section{\label{sec:pert}Gravitational perturbations}

In both the interior (de Sitter) and exterior (Schwarzschild) background
space-times we consider perturbations in the Regge-Wheeler
gauge~\cite{Regge:1957td}, writing
\begin{eqnarray} 
ds^2 &=& ds_0^2 + (\delta_{\rm RW} g_{\mu\nu}) dx^\mu dx^\nu
\end{eqnarray}
with
\begin{widetext}
\begin{equation}
\|\delta_{\rm RW} g_{\mu\nu}\| = 
\left[\begin{array}{cccc}
f(r) H_0(t,r)  Y_{lm} & H_1(t,r) Y_{lm} & \displaystyle 
- h_0(t,r)\frac{1}{\sin\theta}\frac{\partial Y_{lm}}{\partial\varphi} 
&\displaystyle  h_0(t,r){\sin\theta}\frac{\partial Y_{lm}}{\partial\theta} \\ 
*  & \displaystyle \frac{H_2(t,r)  Y_{lm}  }{h(r)} 
& \displaystyle - h_1(t,r)\frac{1}{\sin\theta}\frac{\partial Y_{lm}}
{\partial\varphi} 
&\displaystyle  h_1(t,r){\sin\theta}\frac{\partial Y_{lm}}{\partial\theta} \\
* & * & r^2 K(t,r) Y_{lm} &  0 \\
* & * & * & r^2 \sin^2\theta K(t,r) Y_{lm}  
\end{array}\right]\,,
\end{equation}
\end{widetext}
where $Y_{lm}(\theta,\phi)$ denotes the ordinary spherical harmonics and
``$*$'' stands for terms obtainable by symmetry. In this gauge the
perturbations split into two independent sets: the metric functions $h_0$ and
$h_1$ are \emph{axial} or \emph{odd parity} perturbations, while
$H_0\,,H_1\,,H_2\,,K$ are \emph{polar} or \emph{even parity}
perturbations. The linearized Einstein equations automatically require
$H_0=H_2 \equiv H$.

In the rest of this section we work out perturbations of the gravastar
space-time, including the dynamics of the shell itself, in three steps. In
Sec.~\ref{sec:in}, we present the well-known solution of the perturbation
equations in de Sitter space-time in terms of hypergeometric functions and
choose the solution that is regular at the origin ($r=0$).  In
Sec.~\ref{sec:out} we review metric perturbations in the Schwarzschild
exterior. Finally, in Sec.~\ref{sec:match} we work out the junction conditions
relating the interior and exterior Regge-Wheeler perturbations.

%%%%%%%%%%%%%%%%%%%%%%%%%%%%%%%%%%%%%%%%%%%%%%%%%%%%%%%%%%%%%%%%%%%%%%%%%%%%%%%%%%%%%%%%%%%%%
\subsection{\label{sec:in}The de Sitter interior}
%%%%%%%%%%%%%%%%%%%%%%%%%%%%%%%%%%%%%%%%%%%%%%%%%%%%%%%%%%%%%%%%%%%%%%%%%%%%%%%%%%%%%%%%%%%%%

The usual way to obtain the interior solution for perturbed stars is by direct
integration of the system of ODEs
\cite{Lindblom:1983ps,Detweiler:1985zz,Chandrasekhar:1991fi}.  Integrating a
regular solution from the center would give boundary conditions at the stellar
radius (or, in the case of a gravastar, at the location of the shell).  For
the de Sitter interior ($r<a$) no numerical integrations are required, because
a regular solution of the perturbation equations can be obtained in terms of
hypergeometric functions \cite{Cardoso:2006bv,Berti:2009kk}.
To establish notation we review the basic equations below. Let us express the
metric in terms of a compactness parameter $C\equiv (2M/a)^3$, related to the
parameter $\mu=M/a$ of Ref.~\cite{Chirenti:2007mk} by $C=8\mu^3$. Then we have (assuming $\Sigma=0$)
\be
f(r) =1-\frac{8\pi\rho}{3}r^2=1-\frac{2M}{a^3}r^2\equiv 1-C(r/2M)^2\,.
\ee
In the de Sitter interior {\it both} axial and polar perturbations can be
reduced to the study of the (frequency-domain) master equation
\be
\frac{d^2 \Psi^{\rm in}}{d r_*^2}+
\left [\omega^2-V_{\rm in}(r)\right ]\Psi^{\rm in}=0\,,\quad r<a\,,
\label{eq:masterrstar}
\ee
where 
\be
V_{\rm in}(r)=\frac{l(l+1)}{r^2}\,f(r)\,
\label{eq:potin}
\ee
and we introduced the tortoise coordinate, defined as usual by the condition
$dr/dr_*=f(r)$, which in this case yields
\be
r_* = \sqrt{\frac{3}{8\pi\rho}} \mathrm{arctanh}
\left[\left(\frac{8\pi\rho r^2}{3}\right)^{1/2}\right]\,,\quad r<a\,.
\ee
In terms of $r$, the master equation reads
\be
\frac{\partial^2 \Psi^{\rm in}}{\partial r^2} +
\frac{f'}{f}\frac{\partial \Psi^{\rm in}}{\partial r}
+\frac{\omega^2-V_{\rm in}(r)}{f^2}\Psi^{\rm in}=0\,,
\label{eq:master}
\ee
where a prime denotes a derivative with respect to $r$. Near the origin,
solutions of Eq.~(\ref{eq:master}) behave as $Ar^{l+1}+Br^{-l}$.  By requiring
regularity at the center ($r=0$) we get, up to a multiplicative constant,
\beq
\Psi^{\rm in}\!\!&=&\!\!
r^{l+1}(1-C(r/2M)^2)^{-i\frac{M\omega}{\sqrt{C}}}\nonumber \\&&\!\!\!\!F
\bigg[
\frac{l+2-i\frac{2M\omega}{\sqrt{C}}}{2},
\frac{1+l-i\frac{2M\omega}{\sqrt{C}}}{2},l+\frac{3}{2},\frac{Cr^2}{4M^2}\bigg],\quad\quad
\label{eq:interior_sol}
\eeq
where $F(a,b,c,z)$ is the hypergeometric function
\cite{Abramowitz:1970as}. From $\Psi^{\rm in}$ and its derivative we can get
the axial perturbation variables in the frequency domain
\cite{Zerilli:1971wd}:
\be
h_1=\frac{r}{f}\Psi^{\rm in}\,,\qquad h_0=
-\frac{i }{\omega}\frac{d}{dr_*}\left(r\Psi^{\rm in}\right)\label{axialpsi}\,.
\ee
The polar metric functions $K$ and $H_1$ can be obtained immediately
from:
\beq
K&=&\frac{l(l+1)}{2r}\Psi^{\rm in}+\frac{d\Psi^{\rm in}}{dr_*}\,,\\
H_1&=&\frac{i\omega r}{f}\left(\frac{\Psi^{\rm in}}{r}+
\frac{d\Psi^{\rm in}}{dr_*}\right)\,.
\label{eq:KH1psi}
\eeq
The quantity $H(=H_0=H_2)$ and its derivatives can then be found from the algebraic relation
\beq
&&\left[\frac{l(l+1)}{2} -\frac{1}{f}-\frac{\omega^2 r^2 }{f}\right]K 
+\left[ -i\omega r -  i\frac{l(l+1) C r}{8 M^2\omega}   \right] H_1 \nonumber \\
&&- \frac{(l-1)(l+2)}{2} H
  =0\,.\label{ArelationsdS}
\eeq
This procedure fixes all metric quantities and their derivatives in the
interior.  Here and henceforth in the paper we drop the dependence of $h_0$,
$h_1$, $H$, $H_1$ and $K$ on $\omega$ and $r$.

%%%%%%%%%%%%%%%%%%%%%%%%%%%%%%%%%%%%%%%%%%%%%%%%%%%%%%%%%%%%%%%%%%%%%%%%%%%%%%%%
\subsection{\label{sec:out}The Schwarzschild exterior}
%%%%%%%%%%%%%%%%%%%%%%%%%%%%%%%%%%%%%%%%%%%%%%%%%%%%%%%%%%%%%%%%%%%%%%%%%%%%%%%%

In the Schwarzschild exterior, axial and polar perturbations obey different
master equations \cite{Chandrasekhar:1985kt}.  The determination of the axial
perturbation variables can still be reduced to the solution of the
Regge-Wheeler equation \cite{Regge:1957td}, a Schr\"odinger-like ODE identical
to Eq.~\eqref{eq:master}:
\be
\frac{\partial^2 \Psi^{\rm out}}{\partial r^2} +
\frac{f'}{f}\frac{\partial \Psi^{\rm out}}{\partial r} +
\frac{\omega^2-V_{\rm out}(r)}{f^2}
\Psi^{\rm out}=0\,,\label{masterSch}
\ee
where
\be
V_{\rm out} = f\,\left(\frac{l(l+1)}{r^2}-\frac{6M}{r^3}\right )\,.
\ee
The metric can then be obtained from Eqs.~(\ref{axialpsi}), with $f(r)$ given
by Eq.~(\ref{fr}).  

The perturbed Einstein equations relate the polar variables ($K$, $H$, $H_1$)
via three differential equations:
\beq
\frac{d}{dr}(fH_1)-i\omega (H+K)&=&0\,,\label{eq:ee1}\\
-i\omega H_1+f(H'-K')+f'H&=&0\,,\label{eq:ee2}\\
K'-\frac{H}{r}+\left[\frac{1}{r}-\frac{f'}{2f}\right]K+i\frac{l(l+1)}{2\omega r^2}H_1&=&0\,,
\label{eq:ee3}
\eeq
and an algebraic relation:
\beq
&&\left[\frac{l(l+1)}{2}-1+\frac{rf'}{2}\left(1-\frac{rf'}{2f}\right)-\frac{\omega^2r^2}{f}\right]K\nonumber \\
&+&\left[-i\omega r+i\frac{l(l+1)}{4\omega}f'\right]H_1\nonumber \\
&-&\left[\frac{l(l+1)}{2}-f+\frac{rf'}{2}\right]H 
=0.\quad\quad\quad
\label{eq:alg}
\eeq
Note that if we make the appropriate choice for $f(r)$,
Eqs.~(\ref{eq:ee1})--(\ref{eq:ee3}) and the algebraic relation also apply to
the interior de-Sitter spacetime [cf. Eq.~\eqref{ArelationsdS}].

The Zerilli function $Z^{\rm out}(r)$~\cite{Zerilli:1971wd}, which satisfies a
wave equation, and its spatial derivative are also constructed from $H_1$ and
$K$ as
\beq
Z^{\rm out}&=&\frac{H_1^{\rm out}-A_3K^{\rm out}}{A_2-A_1 A_3}\,,\\
\frac{dZ^{\rm out}}{dr_*}&=&\frac{A_2K^{\rm out}-A_1H_1^{\rm out}}{A_2-A_1A_3}\,,\label{eq:Zout}
\eeq
where
\beq
A_1&=&\frac{6M^2+\lambda/2\left(1+\lambda/2\right)r^2+3/2\lambda Mr} 
{r^2\left(3M+r\lambda/2\right)}\,,\\
A_2&=&\frac{i\omega\left(-3M^2-3/2\lambda\,M\,r+r^2\lambda/2\right)} 
{r\left(3M+r\lambda/2\right) f}\,,\\
A_3&=&i \omega \frac{r}{f}
\eeq
and $\lambda=(l-1)(l+2)$.  The metric perturbations can then be obtained by
integrating the Zerilli equation outwards, starting from $r=a_+$.
It is also useful to recall that (in the exterior) we can switch from the
Zerilli function $Z^{\rm out}(r)$ (\ref{eq:Zout}) to the Regge-Wheeler
function $\Psi^{\rm out}(r)$ by using a differential relation between polar
and axial variables discussed in Chandrasekhar's book
\cite{Chandrasekhar:1985kt}:
\be\label{chandratr}
\alpha_-\Psi^{\rm out}=\eta Z^{\rm out}-\frac{dZ^{\rm out}}{dr_*}\,,\qquad \frac{d\Psi^{\rm out}}{dr_*}=\alpha_+Z^{\rm out}-\eta\Psi^{\rm out}\,,\nonumber
\ee
where
\be 
\alpha_\pm=\frac{\lambda(\lambda+2)}{12M}\pm i\omega\,,\quad\eta=\frac{\lambda(\lambda+2)}{12M}+\frac{6Mf(r)}{r(\lambda\,r+6M)}\,.\nonumber
\ee
From $Z^{\rm out}$ and $dZ^{\rm out}/dr_*$ we can easily compute $\Psi^{\rm
  out}$ and $d\Psi^{\rm out}/dr_*$ outside the shell and use them as initial
conditions to integrate the Regge-Wheeler equation outwards. Leins {\it et
  al.} \cite{Leins:1993zz} showed that this procedure is convenient to compute
polar oscillation modes by the continued fraction method; more details on the
QNM calculation are given in Appendix \ref{app:cf}.

%%%%%%%%%%%%%%%%%%%%%%%%%%%%%%%%%%%%%%%%%%%%%%%%%%%%%%%%%%%%%%%%%%%%%%%%%%%%%%%%
\subsection{\label{sec:match}The matching conditions}
%%%%%%%%%%%%%%%%%%%%%%%%%%%%%%%%%%%%%%%%%%%%%%%%%%%%%%%%%%%%%%%%%%%%%%%%%%%%%%%%

In this section we discuss the most delicate part of the perturbation problem,
namely the junction conditions for the Regge-Wheeler perturbation variables
across the shell. Here we only outline our strategy and present the results;
more details are given in Appendix \ref{app:match}.  Technically, the
application of Israel's junction conditions is easier if the shell's world
tube happens to coincide with a fixed coordinate sphere at constant radius.
However this is incompatible with choosing the Regge-Wheeler gauge in both the
interior and the exterior, which is convenient to cast the perturbation
equations into simple forms.  In fact, such a choice of gauge does not leave
any freedom. We must explicitly parametrize the 3-dimensional motion of each
mass element on the shell and then perform the matching on a moving shell.
In order to take advantage of both the simplicity of the field equations and
the convenience of matching on a fixed coordinate sphere, we carry out the
matching in the following way.  We first construct a particular coordinate
transformation (for both the exterior and interior space-times) such that in
the new coordinate system, any mass on the shell remains static on the
coordinate sphere with $r=a$.  In this new coordinate system the metric
perturbations will no longer be Regge-Wheeler, but will be augmented by
quantities that carry information about how masses on the shell move in the
Regge-Wheeler gauge. The stress-energy tensor of masses on the shell will
correspondingly be modified. We then carry out the matching at $r=a$ and
obtain junction conditions relating the interior and exterior metric
perturbations, plus equations of motion for matter on the shell.  As could be
anticipated from the general features of oscillations of nonrotating stars,
axial perturbations do not couple to matter motion and the axial junction
conditions are very simple, basically imposing continuity of the master
variable $\Psi$ and of its first derivative.  Polar perturbations, on the
other hand, do couple to matter motion, so polar junction conditions do
involve the shell dynamics, i.e. its equation of state.

We parametrize the world line of matter elements on the shell in terms of the
coordinates $(t,r,\theta,\varphi)$ as follows:
\be
\left\{
\begin{array}{llll}
\displaystyle 
t =\tau/\sqrt{f(a)} +  \delta t(\tau,\theta_*,\varphi_*)\,, 
\\
\displaystyle 
r =a +\delta r (\tau,\theta_*,\varphi_*)\,, \\
\displaystyle 
\theta =\theta_* +\delta \theta (\tau,\theta_*,\varphi_*)\,, \\
\displaystyle 
\varphi =\varphi_* +\delta \varphi (\tau,\theta_*,\varphi_*)\,,
\end{array}
\right.\label{shellpos}
\ee
where $\theta_*$ and $\varphi_*$ identify physical mass elements on the
sphere, while $\tau$ parametrizes their proper time.  Note that the Lagrangian
equations of motion will {\it not} be the same for the interior and exterior
space-times. Therefore, points with the same $t\,,\theta$ and $\varphi$
coordinates are not in general the same when viewed from the interior and from
the exterior.
As shown in Appendix \ref{app:match}, the four-velocity of the mass element
$(\theta_*,\varphi_*)$ at the scaled proper time $\hat{t} \equiv
\tau/\sqrt{f(a)}$ is, to leading order in the perturbation variables:
\begin{equation}
u^\alpha = \left[f(a)\right]^{-1/2}
(1+\delta\dot{t},\delta\dot{r},\delta\dot{\theta},\delta\dot{\varphi})\,.
\label{4vel}
\end{equation}
We now carry out a gauge transformation which maps the shell to a fixed
location (note that two different gauge transformations are required for the
exterior and for the interior). For any general gauge transformation $\bar
x^{\bar\alpha} = x^\alpha-\xi^{\alpha}(x^\mu)$ we have, to first order in
$\xi^\mu$,
\begin{equation}
\delta g_{\alpha\beta}=\bar g_{\alpha\beta} (\bar x^\mu)-
g_{\alpha\beta} (\bar x^\mu) = \xi_{\beta;\alpha}(\bar x^\mu)+
\xi_{\alpha;\beta}(\bar x^\mu)\,,\label{deltagmunu}
\end{equation}
where the semicolon represents a covariant derivative with respect to the
four-metric and $\bar g_{\alpha\beta} (\bar x^\mu)$ is the metric tensor in
the new coordinate system. We impose that, when evaluated at
$(t,r,\theta,\varphi) = (\hat t, a,\theta_*,\varphi_*)$, the vector
$\xi^\alpha$ coincides with $(\delta t,\delta r,\delta\theta,\delta\varphi)$,
so that in the new coordinate system we will have
\be
\left\{
\begin{array}{llll}
\displaystyle 
\bar t &=&\tau/\sqrt{f(a)}\,,  \\
\displaystyle 
\bar r &=& a\,,\\
\displaystyle 
\bar \theta &=& \theta_*\,, \\
\displaystyle 
\bar \varphi  &=&\varphi_*\,,
\end{array}
\right.\label{shellpositionbar}
\ee
where we are ignoring second-order corrections. The full metric in the new
coordinate system is
\begin{equation}
\bar g_{\alpha\beta} = g^{\rm (0)}_{\alpha\beta} + \delta_{\rm RW} g _{\alpha\beta}
+ \delta g _{\alpha\beta}\,,\label{gbarmunu}
\end{equation}   
where $g^{\rm (0)}_{\alpha\beta}$ is the static gravastar background metric,
given by Eq.~(\ref{eq:g0}). The explicit form of $\xi^\mu$ and the
corresponding changes in the metric components, Eq.~(\ref{deltagmunu}), are
given in Appendix \ref{app:match}, where the equations of motion, as well as
the gauge transformation, are presented systematically in a multipole
expansion. We then match components of $\bar g_{\alpha\beta}$ along the shell,
which now simply sits at $(\hat t,a,\theta_*,\varphi_*)$, and apply Israel's
junction conditions to the extrinsic curvature given by $\bar
g_{\alpha\beta}$. For axial perturbations these matching conditions read
\be 
\LL h_0\RR=0\,,\quad \LL \sqrt{h}h_1\RR=0\,.
\label{junctionaxial}
\ee
For thin shell gravastars $h(r)$ is continuous across the shell, implying
continuity of the Regge-Wheeler function $\Psi$ and its derivative $\Psi'$
across the shell [cf. Eq.~(\ref{axialpsi})].  In more general cases where
$h(r)$ may have a discontinuity across the shell the axial junction conditions
(\ref{junctionaxial}) show that $\Psi$ must also be discontinuous.

The treatment of polar perturbations is more involved and it yields the
following relations, determining the ``jump'' of the polar metric functions at
the shell:
\beq 
\label{junctionpolar1}
&&[[K]]=0\,,\\
\label{junctionpolar2}
&&[[K']]=-8\pi\frac{\delta\Sigma}{\sqrt{f(a)}}\,,\\
&&\frac{2M}{a^2}\LL H\RR-[[H\,f']]-2f(a)[[H']]+4i\omega[[H_1]]\nonumber \\
&=&16\pi\sqrt{f(a)}(1+2v_s^2)\delta\Sigma
\label{junctionpolar3}\,.
\eeq
Note that $f(r)$ is continuous across the shell [cf. Eq.~\eqref{fcont}].  The
parameter $v_s$ in the equations above depends on the equation of state on the
thin shell, $\Theta=\Theta(\Sigma)$:
\be 
v_s^2\equiv-\left(
\frac{\partial\Theta}{\partial\Sigma}
\right)_{\Sigma=0}\,,\label{eq:v}
\ee
and it has the dimensions of a velocity.
One might naively interpret $v_s$ as the speed of sound on the thin shell and
require $v_s\leq1$, i.e. that the speed of sound cannot exceed the speed of
light. Furthermore, for a shell of ordinary, stable matter we would have
$v_s^2>0$. The standard argument used to deduce that $v_s^2>0$ does not
necessarily hold when one deals with exotic matter (as in the case of
gravastars and wormholes), so the specification of upper and lower bounds on
$v_s$ may require a detailed microphysical model of the exotic matter itself
\cite{Visser:1995cc,Poisson:1995sv}. In our discussion of gravastar stability
we will consider the whole range of $v_s$, but we will primarily focus on the
range $0\leq v_s^2\leq 1$.  The application of the polar junction conditions
is more involved than the axial case due to their complexity, which arises
from the fact that polar perturbations couple to oscillations of the shell.

Here we note that even though we have three quantities ($K$, $H$, $H_1$) that
satisfy a coupled system of first-order ODEs in both the interior and
exterior, in each region there is an algebraic relation relating the three
quantities. For this reason we only need to impose two independent junction
conditions, and Eqs.~\eqref{junctionpolar1}--\eqref{junctionpolar3} provide
exactly two independent relations among $K$, $H$ and $H_1$ (after eliminating
$\delta \Sigma$).  Alternatively, the number of junction conditions can be
obtained considering that all metric perturbations can be expressed in terms
of $\Psi^{\rm in}$ and $\Psi^{\rm out}$, and that each of these master
variables satisfies a second-order ODE.  More specifically, we use
Eqs.~\eqref{junctionpolar1}--\eqref{junctionpolar3} to determine two relations
among $(\Psi^{\rm in},\partial_r \Psi^{\rm in})$ and $(\Psi^{\rm
  out},\partial_r \Psi^{\rm out})$, plus the corresponding $\delta \Sigma$.
\begin{center}
\begin{figure*}[ht]
\begin{tabular}{cc}
\includegraphics[width=8.8cm,clip=true]{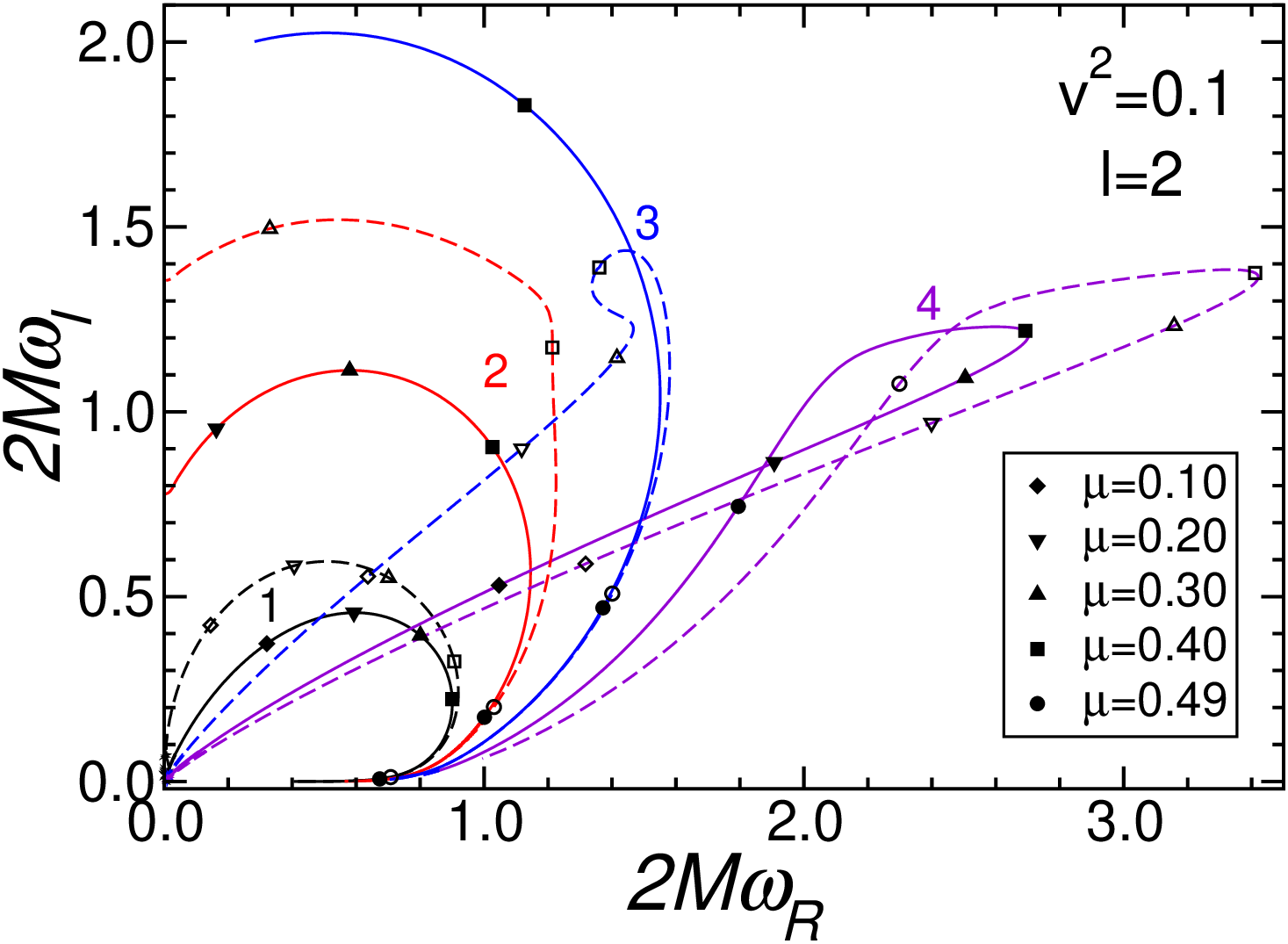}&
\includegraphics[width=8.8cm,clip=true]{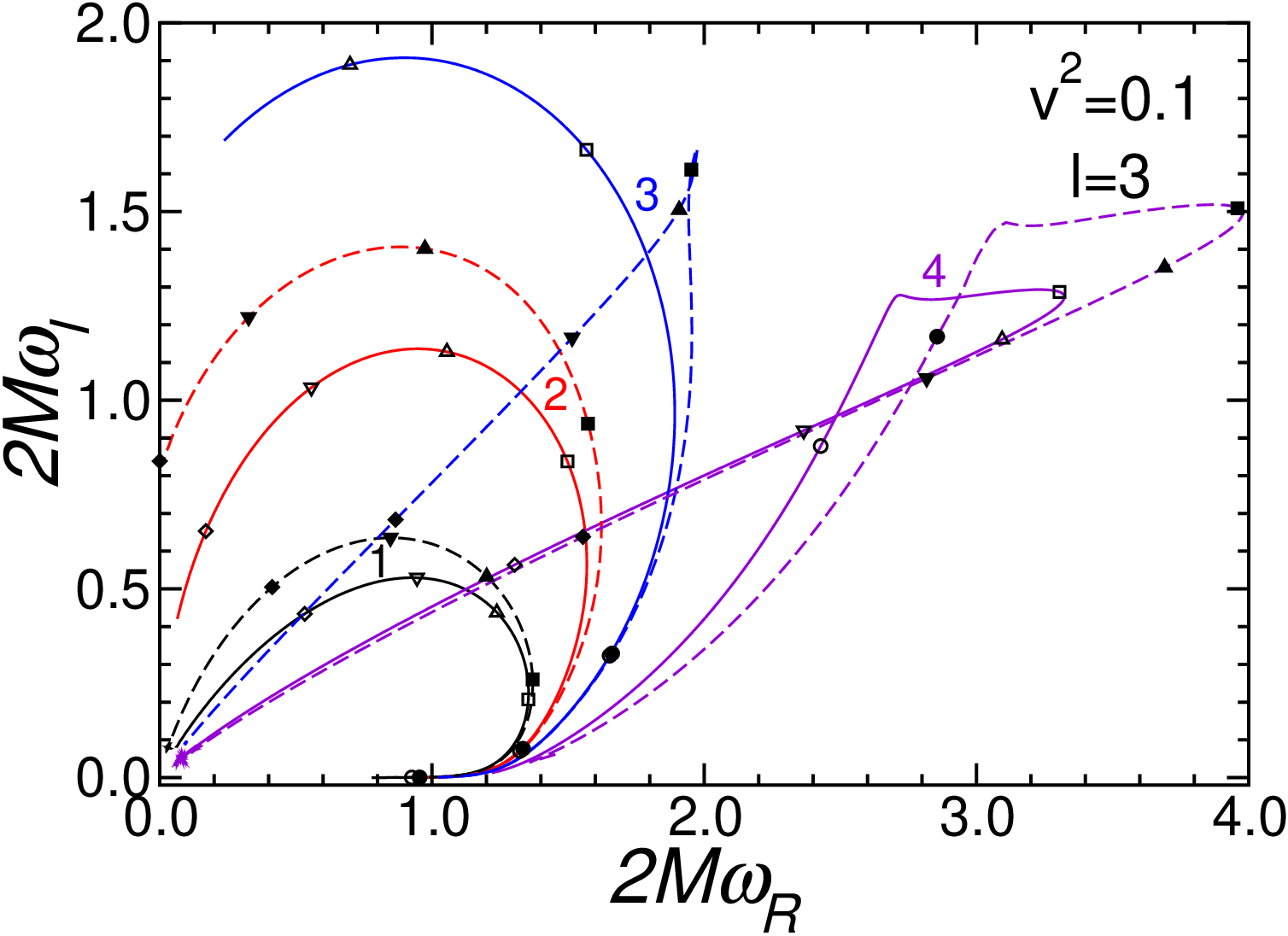}
\end{tabular}
\caption{First few axial (continuous lines) and polar (dashed lines) QNMs of a
  thin-shell gravastar with $v_s^2=0.1$. In the left panel we follow modes
  with $l=2$ as the compactness $\mu$ varies. In the right panel we do the
  same for modes with $l=3$. Along each track we mark by different symbols (as
  indicated in the legend) the points where $\mu=0.1\,,0.2\,,0.3\,,0.4$ and
  $0.49$. Our numerical method becomes less reliable when $2M\omega_I$ is
  large and when the modes approach the pure-imaginary axis. Numbers next to
  the polar and axial modes refer to the overtone index $N$ ($N=1$ being the
  fundamental mode).}
\label{fig:axpolQNMs}
\end{figure*}
\end{center}
\begin{center}
\begin{figure*}[ht]
\begin{tabular}{cc}
\includegraphics[width=8.8cm,clip=true]{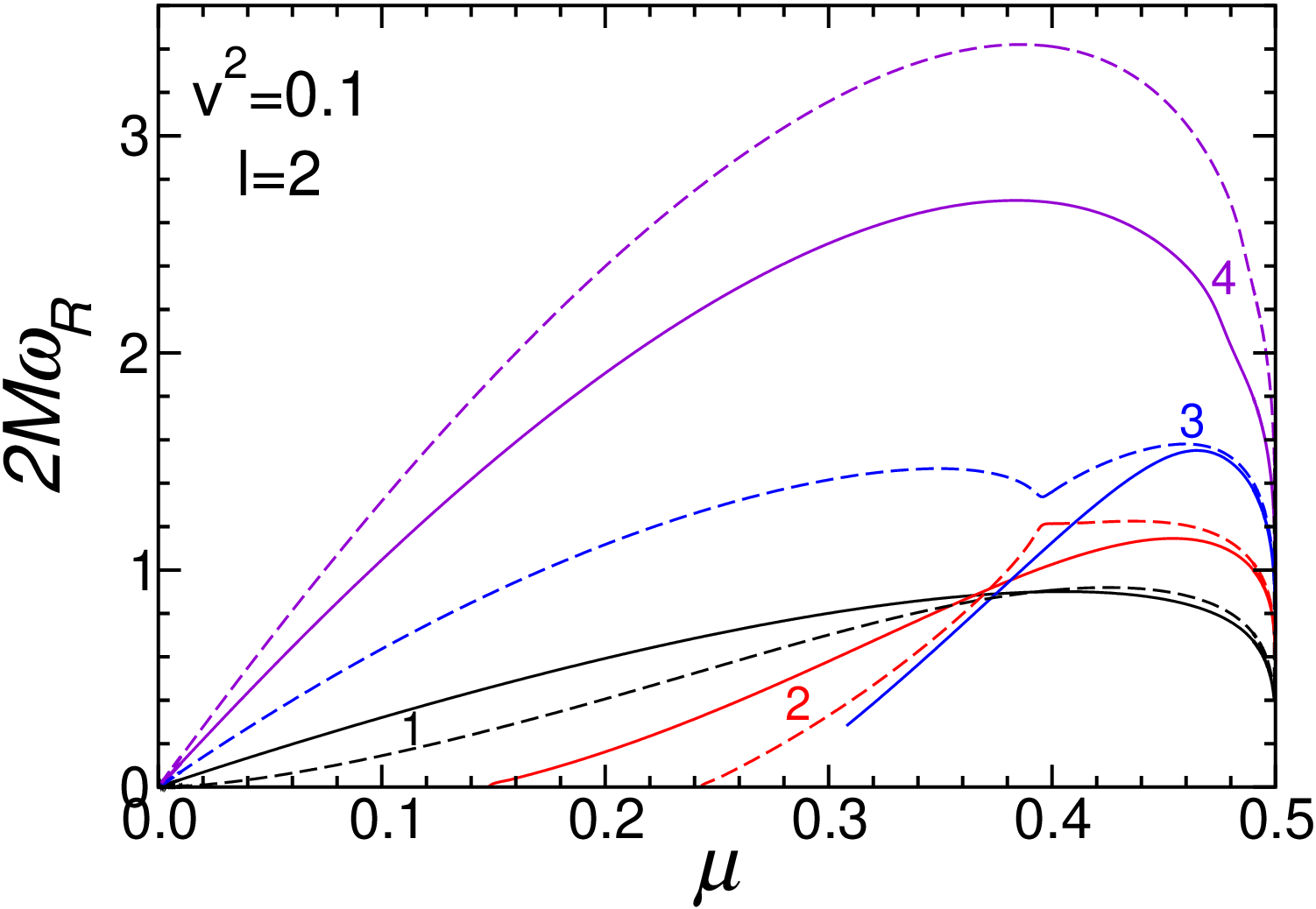}&
\includegraphics[width=8.8cm,clip=true]{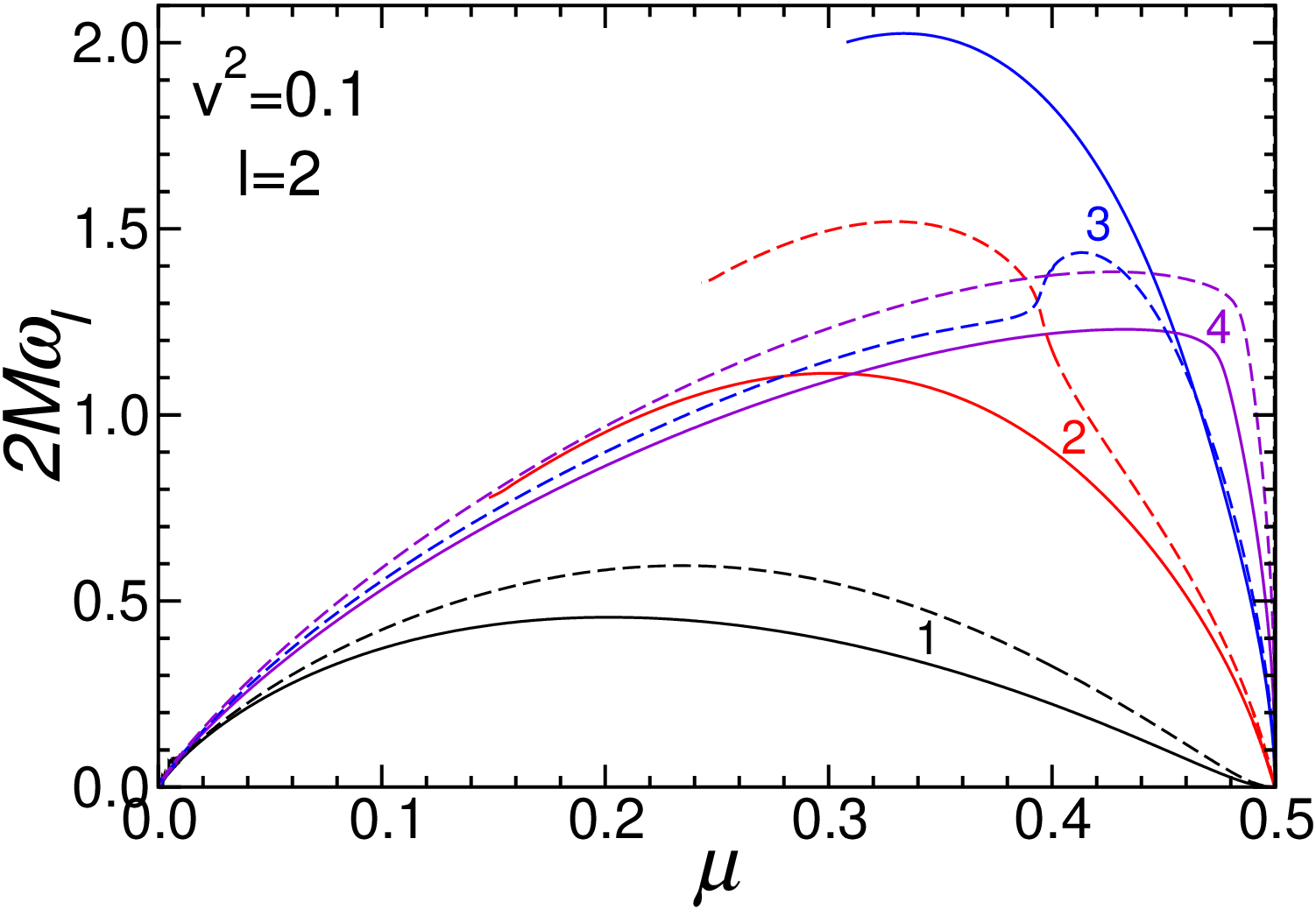}
\end{tabular}
\caption{Real (left) and imaginary (right) part of polar and axial QNMs with
  $l=2$ as functions of $\mu$ for $v_s^2=0.1$. Linestyles are the same as in
  Fig.~\ref{fig:axpolQNMs}. Numbers refer to the overtone index.}
\label{fig:modes}
\end{figure*}
\end{center}
\begin{center}
\begin{figure*}[ht]
\begin{tabular}{cc}
\includegraphics[width=8.8cm,clip=true]{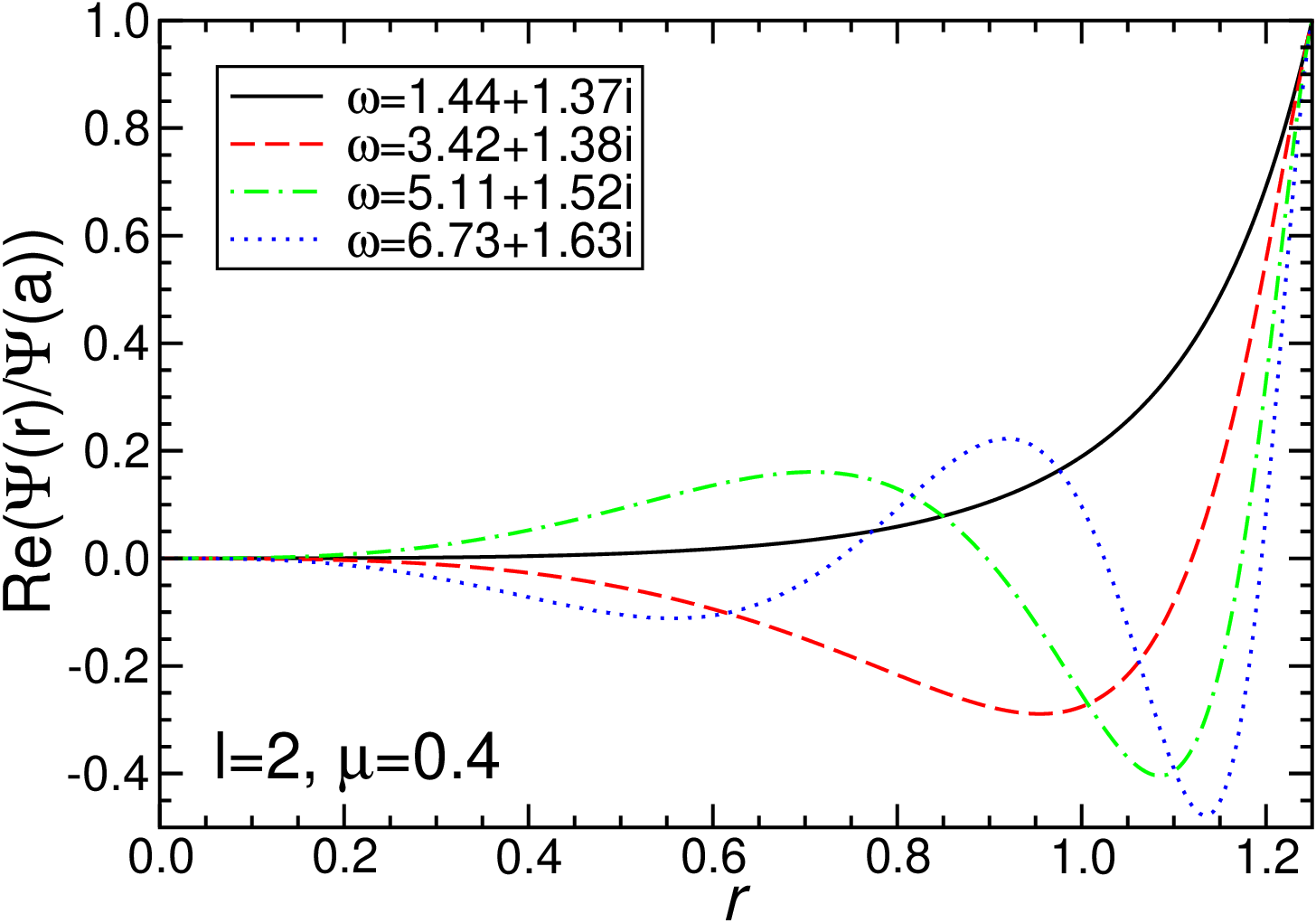}&
\includegraphics[width=8.8cm,clip=true]{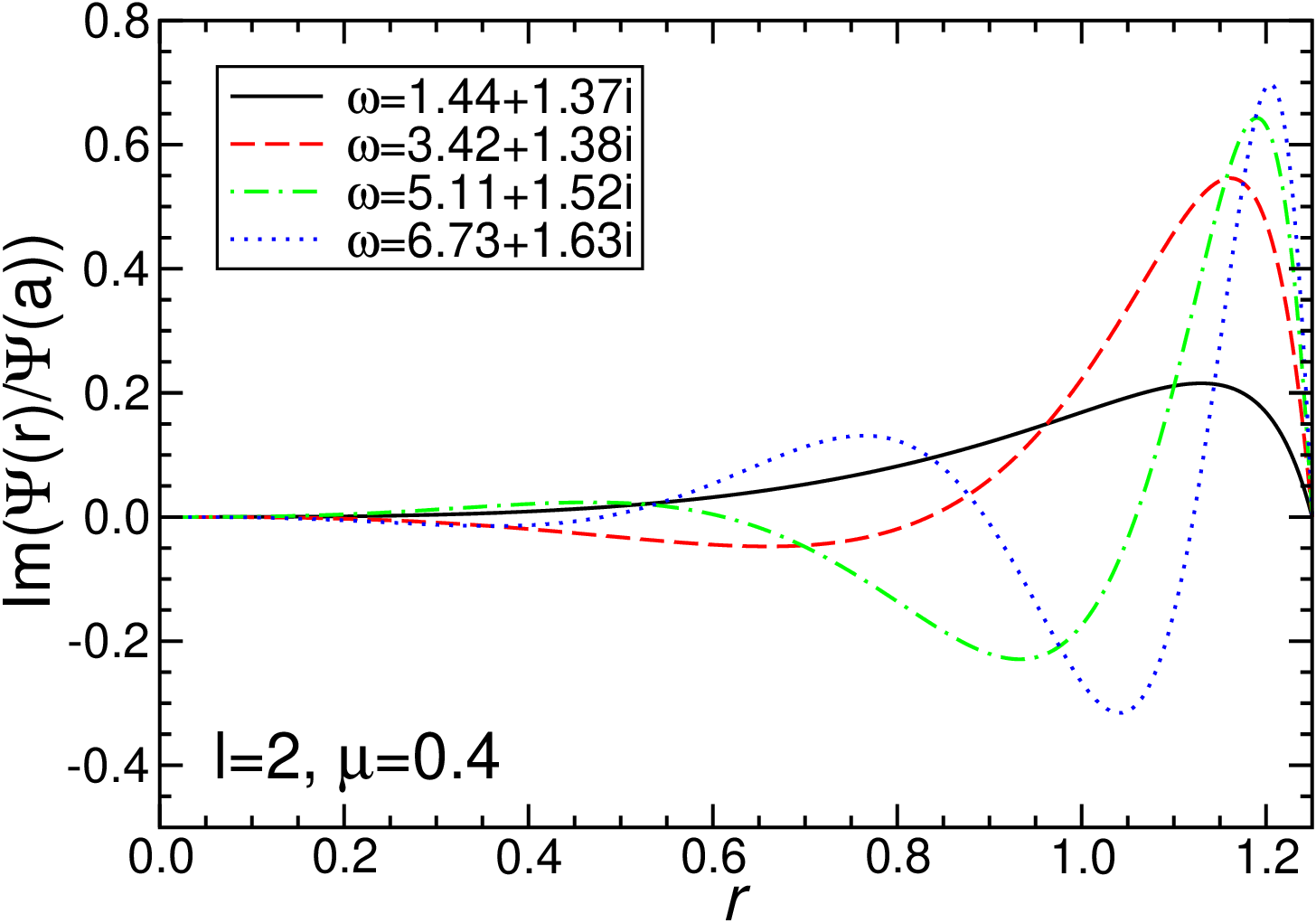}\\
\includegraphics[width=8.8cm,clip=true]{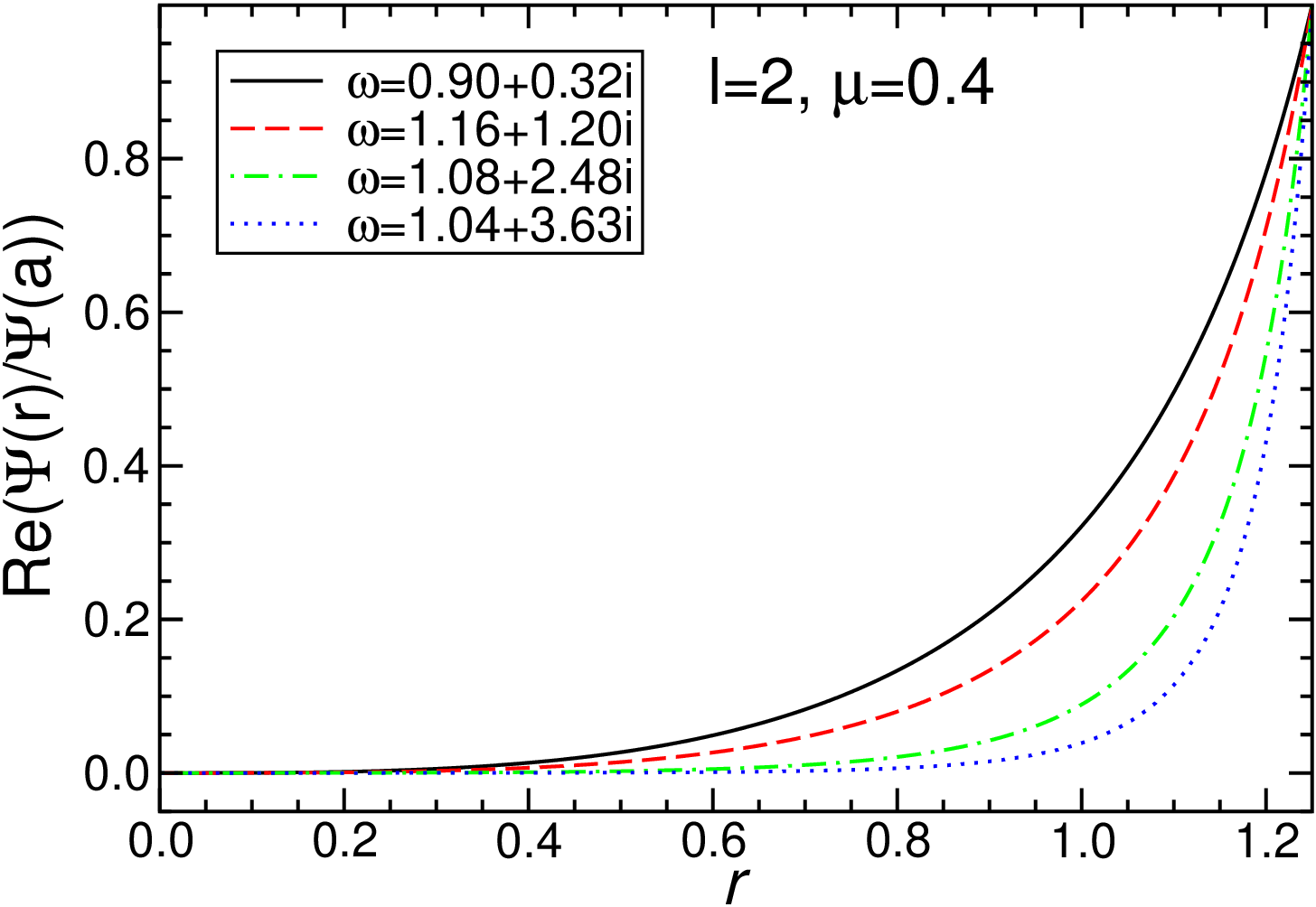}&
\includegraphics[width=8.8cm,clip=true]{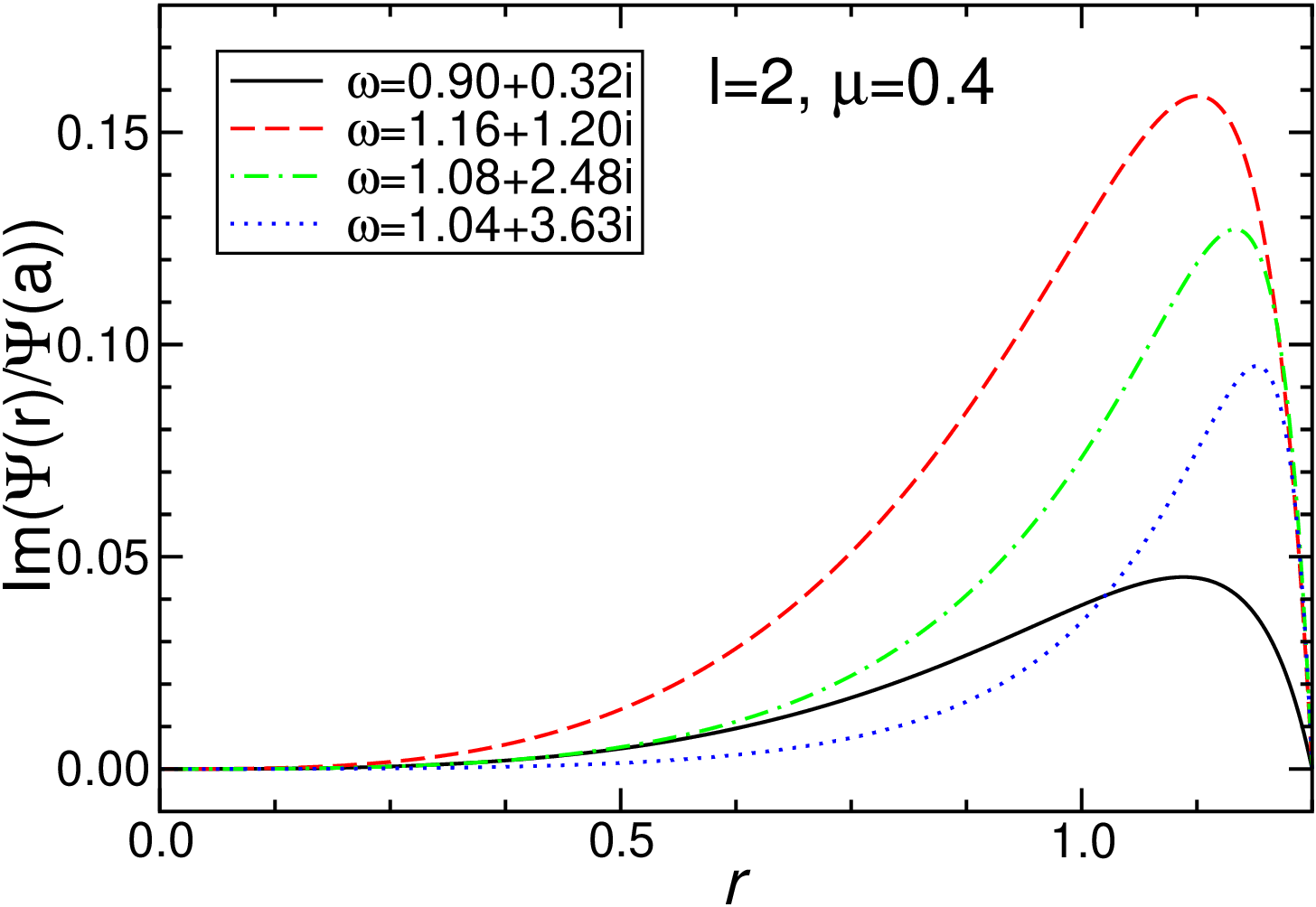}\\
\end{tabular}
\caption{Top row: real and imaginary part of the wavefunction in the interior
  for the first four $w$-modes. Bottom row: real and imaginary part of the
  wavefunction in the interior for the first four $w_{\rm II}$-modes. In both
  cases we consider polar QNMs with $l=2$ and $v_s^2=0.1$.}
\label{fig:smodes}
\end{figure*}
\end{center}
%
%%%%%%%%%%%%%%%%%%%%%%%%%%%%%%%%%%%%%%%%%%%%%%%%%%%%%%%%%%%%%%%%%%%%%%%%%%%%%%%%
\section{\label{sec:results}Numerical Results}
%%%%%%%%%%%%%%%%%%%%%%%%%%%%%%%%%%%%%%%%%%%%%%%%%%%%%%%%%%%%%%%%%%%%%%%%%%%%%%%%

Some axial and polar QNM frequencies for a static thin-shell gravastar, as
computed by the continued fraction method, are plotted in
Figs.~\ref{fig:axpolQNMs} and \ref{fig:modes}.  The C$++$ code used for the
calculations is an update of the Fortran code used in
Ref.~\cite{Benhar:1998au} to verify and extend results on stellar oscillations
by Kokkotas \cite{Kokkotas:1994an} and Leins, Nollert and Soffel
\cite{Leins:1993zz}. For axial modes, our numerical results are in excellent
agreement with the thin-shell limit of the QNM frequencies computed by
Chirenti and Rezzolla \cite{Chirenti:2007mk} and with Fiziev's calculation of
the axial QNMs of ultracompact objects with a totally reflecting surface
(compare figures 3 and 4 of Ref.~\cite{Fiziev:2005ki}).

To find the QNM frequencies we adopt the following numerical procedure. We
usually fix $\mu=0.4$ and (for polar perturbations) we choose a constant value
of $v_s$. In the calculations leading to Figs.~\ref{fig:axpolQNMs} and
\ref{fig:modes} we chose, somewhat arbitrarily, $v_s^2=0.1$. Later in this
section we will discuss the dependence of polar modes on $v_s$.

As explained in Appendix \ref{app:cf}, within the continued fraction method
the complex QNM frequencies can be determined as the roots of any of the $n$
equations $f_n(\omega)=0$ [cf. Eq.~(\ref{cuore2})], where $n$ is the
``inversion index'' of the continued fraction. To locate QNMs we first fix a
value of $\mu$ (usually $\mu=0.4$). We compute the real and imaginary parts of
$f_n(\omega)$ for a given inversion index $n$ on a suitably chosen grid of
$(\omega_R,\omega_I)$ values, and make contour plots of the curves along which
the two functions are zero. The intersections of these curves are used as
initial guesses for the quasinormal frequencies; more precise values are then
obtained using Muller's method \cite{NRCPP}. For fixed $n$ (say, $n=0$) this
method singles out some spurious roots besides the physical QNM
frequencies. The spurious roots can easily be ruled out, since they are not
present for different values of $n$.  Looking for roots with $n=0$ is usually
sufficient, but sometimes we get more stable numerical solutions for $n=1$ and
$n=2$ when the QNMs have large imaginary part ($\omega_I \gtrsim 1.5$ or so).

The QNM spectrum with $\mu=0.4$ corresponds to the empty (polar) and filled
(axial) squares in the left panel of Fig.~\ref{fig:axpolQNMs}, respectively.
Starting from $\mu=0.4$, we follow each QNM as $\mu\to 0$ and as $\mu\to 1/2$
to produce the tracks displayed in the figure. For this value of $v_s$ some of
these tracks end at the origin in the limit $\mu\to 0$, while others hit the
pure-imaginary axis at some finite limiting compactness $\mu_{\rm imag}$
(e.g. $\mu_{\rm imag}\simeq 0.24$ for the first polar mode).  The mode
frequencies usually move clockwise in the complex plane (with the exception of
QNMs displaying ``loops'') as $\mu$ is increased. The imaginary part of both
axial modes (continuous lines) and polar modes (dashed lines) becomes very
small as $\mu\to 1/2$, i.e. when the gravastar most closely approximates a
BH. The behavior is perhaps clearer from Fig.~\ref{fig:modes}, where we
separately show the real and imaginary parts as functions of $\mu$.

For both axial and polar spectra the dependence of the mode frequencies on the
gravastar compactness resembles that of ``ordinary'' ultracompact stars: see
e.g. Fig.~3 of Ref.~\cite{Andersson:1995ez}. Intuitive models that capture
most of the physics of this problem have been presented in
Refs.~\cite{Kokkotas:1986gd,Andersson:1996yt}. In their terminology, modes
that emerge from the origin in Fig.~\ref{fig:axpolQNMs} when $\mu\sim0$ are
$w$-modes or {\em curvature} modes, roughly corresponding to waves trapped
inside the star. Modes emerging from the imaginary axis at some generally
nonvanishing compactness are $w_{\rm II}$-modes \cite{Leins:1993zz} or {\it
  interface} modes, and they are similar in nature to acoustic waves scattered
off a hard sphere \cite{Andersson:1996yt,Berti:2002zz}. The only qualitative
difference with Fig.~3 of Ref.~\cite{Andersson:1995ez} are the ``loops''
appearing for higher-order $w$-modes, for which we have no analytical
understanding.

The fact that $w$-modes are effectively waves ``trapped inside the star'',
while $w_{\rm II}$-modes are {\it interface} modes, similar to acoustic waves
scattered off a hard sphere, is also clear from the behavior of their
wavefunctions in the stellar interior. The real and imaginary parts of the
wavefunctions are shown in Fig.~\ref{fig:smodes} (see Ref.~\cite{Leins:1993zz}
for comparison with the wavefunctions of ordinary stars). This figure shows
eigenfunctions computed at the polar QNM frequencies for the first {\it four}
$w$-modes and $w_{\rm II}$-modes with $l=2$ and $v_s^2=0.1$. The plot shows
that $w$-modes can be thought of as standing waves inside the gravastar, and
that the overtone number corresponds to the number of nodes in the real and
imaginary parts of the eigenfunction. The situation is different for $w_{\rm
  II}$-modes, where the wavefunction has a maximum close to the shell, as
expected for interface modes.
%The eigenfunctions of weakly damped modes are just like those of wII modes!

One of our most important conclusions is that {\em neither axial nor polar
  modes of a gravastar reduce to the QNMs of a Schwarzschild BH when $\mu\to
  1/2$}. In this limit, the real part of most modes is extremely small (much
smaller than the Schwarzschild result, $2M\omega_R\simeq 0.74734$ for the
fundamental mode with $l=2$ \cite{Berti:2009kk}).  Indeed, the QNM spectrum is
drastically different from the QNM spectrum of a Schwarzschild BH: when
$\mu\to 1/2$ the entire spectrum seems to collapse towards the origin. This is
in sharp contrast with the Schwarzschild BH case and, as first noted in
Ref.~\cite{Chirenti:2007mk}, it can be used to discern a very compact
gravastar from a BH.  In Appendix \ref{app:highC} we prove analytically that
the QNM frequencies of a gravastar do not reduce to those of a Schwarzschild
BH as $\mu\to 1/2$. The proof is based on the observation that the Zerilli
wavefunction for polar modes is continuous in this limit.

\begin{center}
\begin{figure*}[ht]
\begin{tabular}{cc}
\includegraphics[width=8.8cm,clip=true]{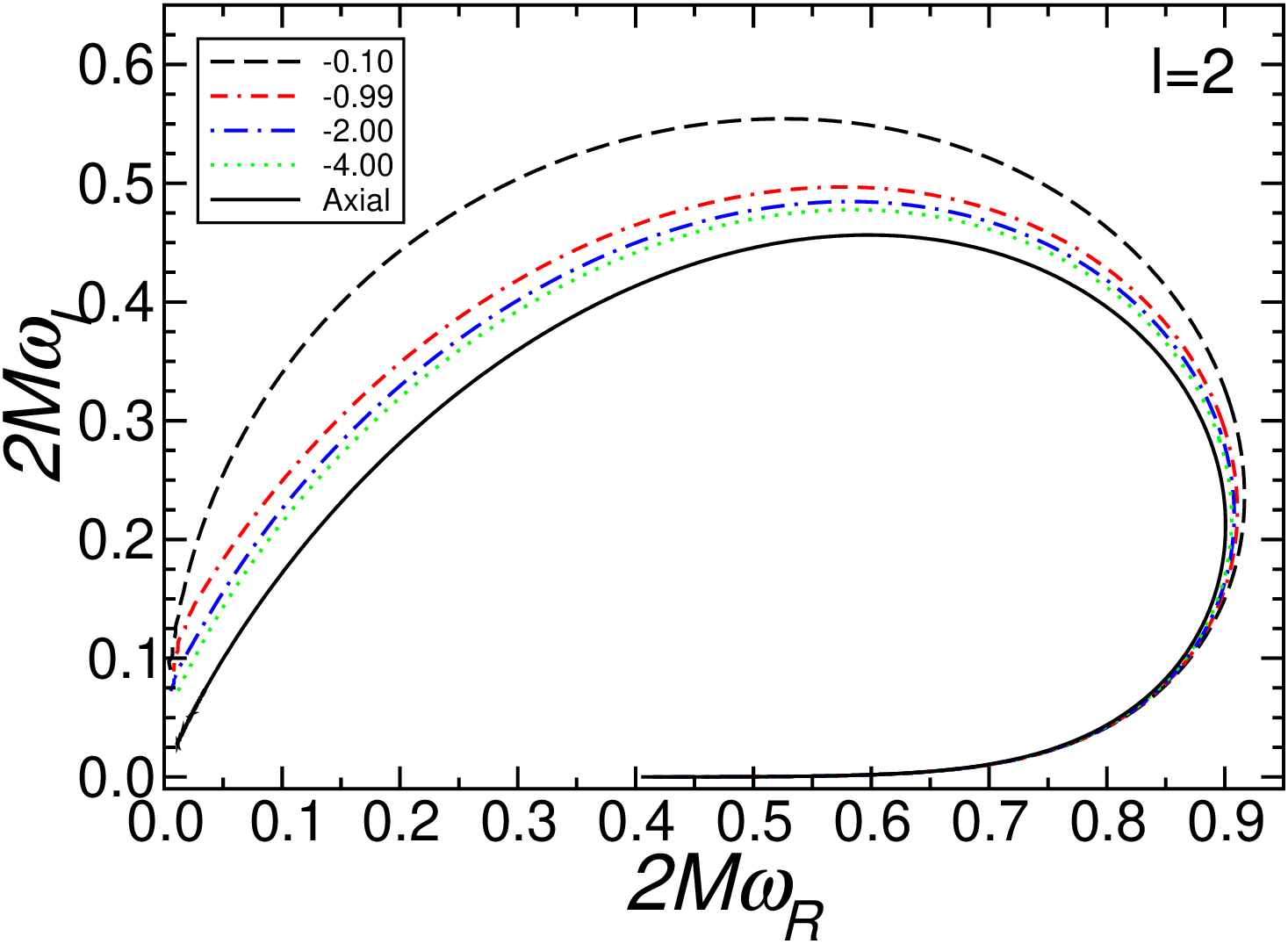}&
\includegraphics[width=8.8cm,clip=true]{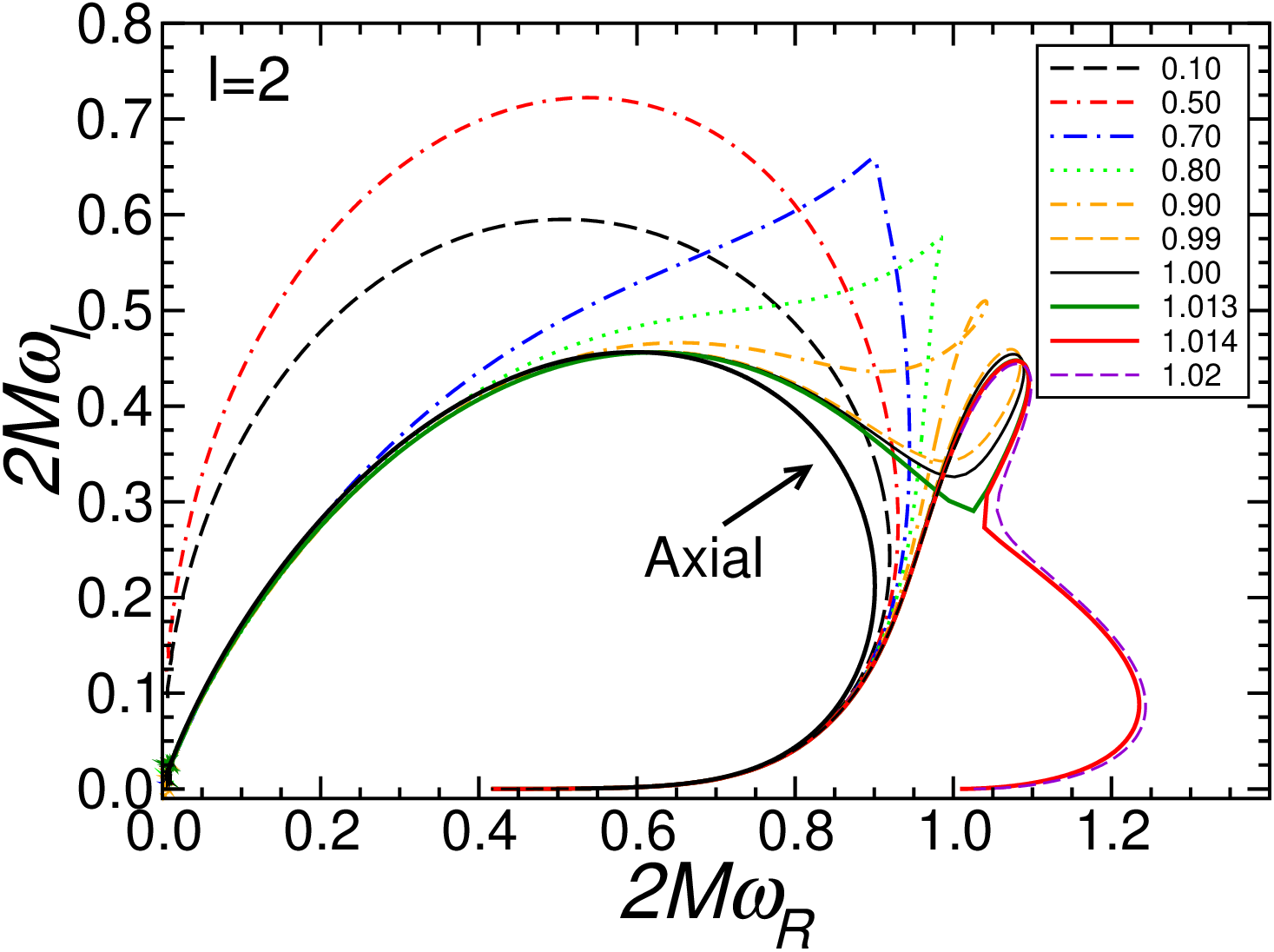}\\
\end{tabular}
\caption{Tracks of the fundamental polar and axial $w$-modes for different
  values of the ``sound speed'' parameter $v_s$ when $v_s^2<0$ (left) and when
  $v_s^2>0$ (right). Different linestyles correspond to different values of
  $v_s^2$, as indicated in the legend.}
\label{fig:polarQNMsv}
\end{figure*}
\end{center}

It is clear from the figures that axial and polar modes do {\it not} have the
same spectra for general values of $\mu$. However, figures
\ref{fig:axpolQNMs}, \ref{fig:modes} and \ref{fig:polarQNMsv} provide evidence
that axial and polar QNMs do become isospectral when the gravastar compactness
approaches that of a Schwarzschild BH ($\mu\to 0.5$). An analytic proof of
isospectrality in the high-compactness limit is given in Appendix
\ref{app:highC}, where we show that in this limit both the Zerilli and
Regge-Wheeler functions are continuous at the shell.

From the matching conditions (\ref{junctionaxial}), (\ref{junctionpolar1}),
(\ref{junctionpolar2}) and (\ref{junctionpolar3}) it is quite clear that polar
QNMs (unlike axial QNMs) should depend on $v_s$, i.e. [by Eq.~(\ref{eq:v})] on
the equation of state on the shell. This is a new feature that does not arise
in the case of axial perturbations \cite{Chirenti:2007mk}. The situation
closely parallels the ordinary stellar perturbation problem
\cite{Chandrasekhar:1991fi,ChandrasekharFerrari}. The role played by the
equation of state in the dynamical stability of gravastars against {\it
  spherically symmetric} perturbations was discussed in
Ref.~\cite{Visser:2003ge}. Our calculations extend the considerations of that
paper to nonradial oscillations.

\begin{center}
\begin{figure*}[ht]
\begin{tabular}{cc}
\includegraphics[width=8.8cm,clip=true]{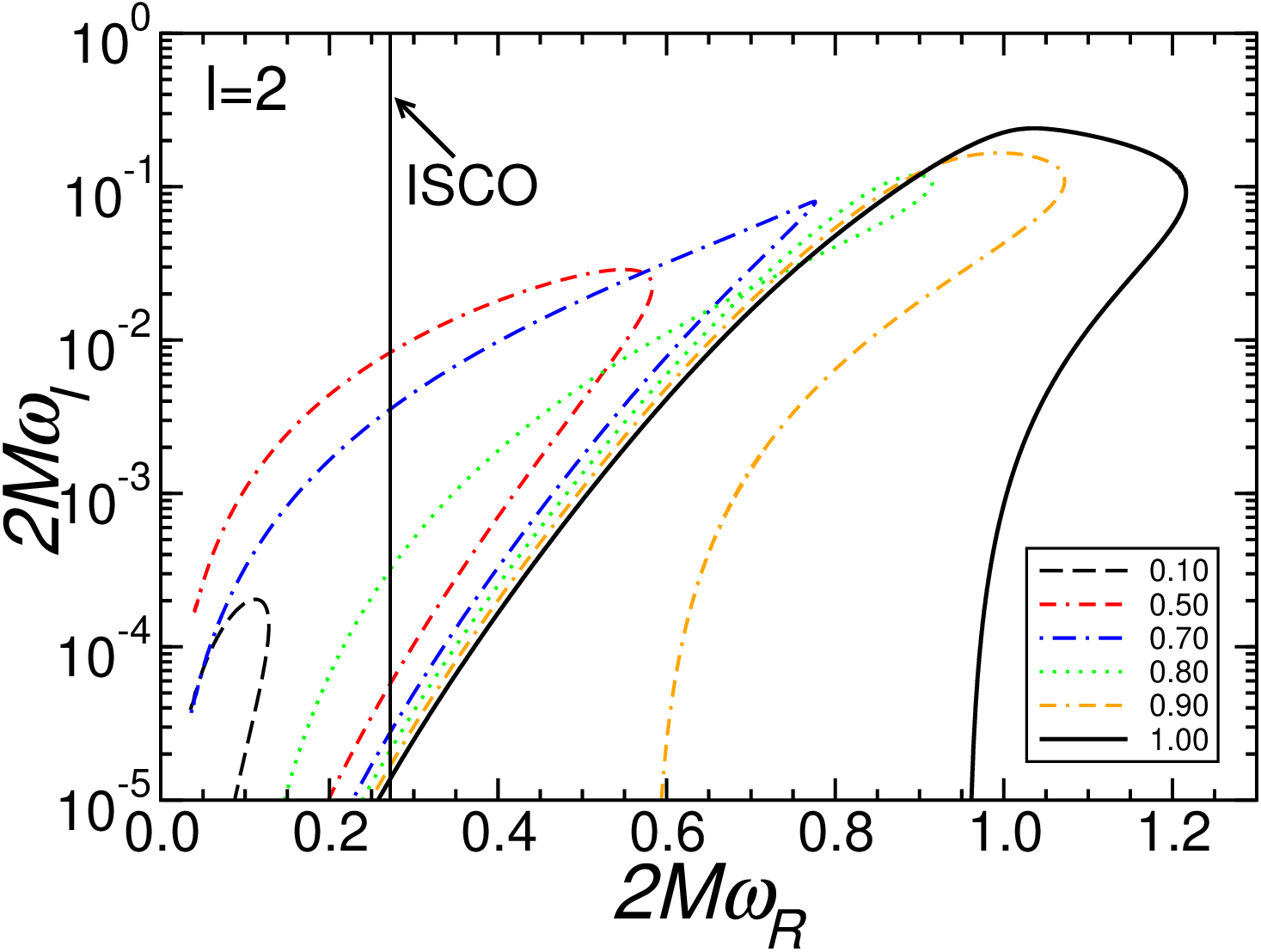}&
\includegraphics[width=8.8cm,clip=true]{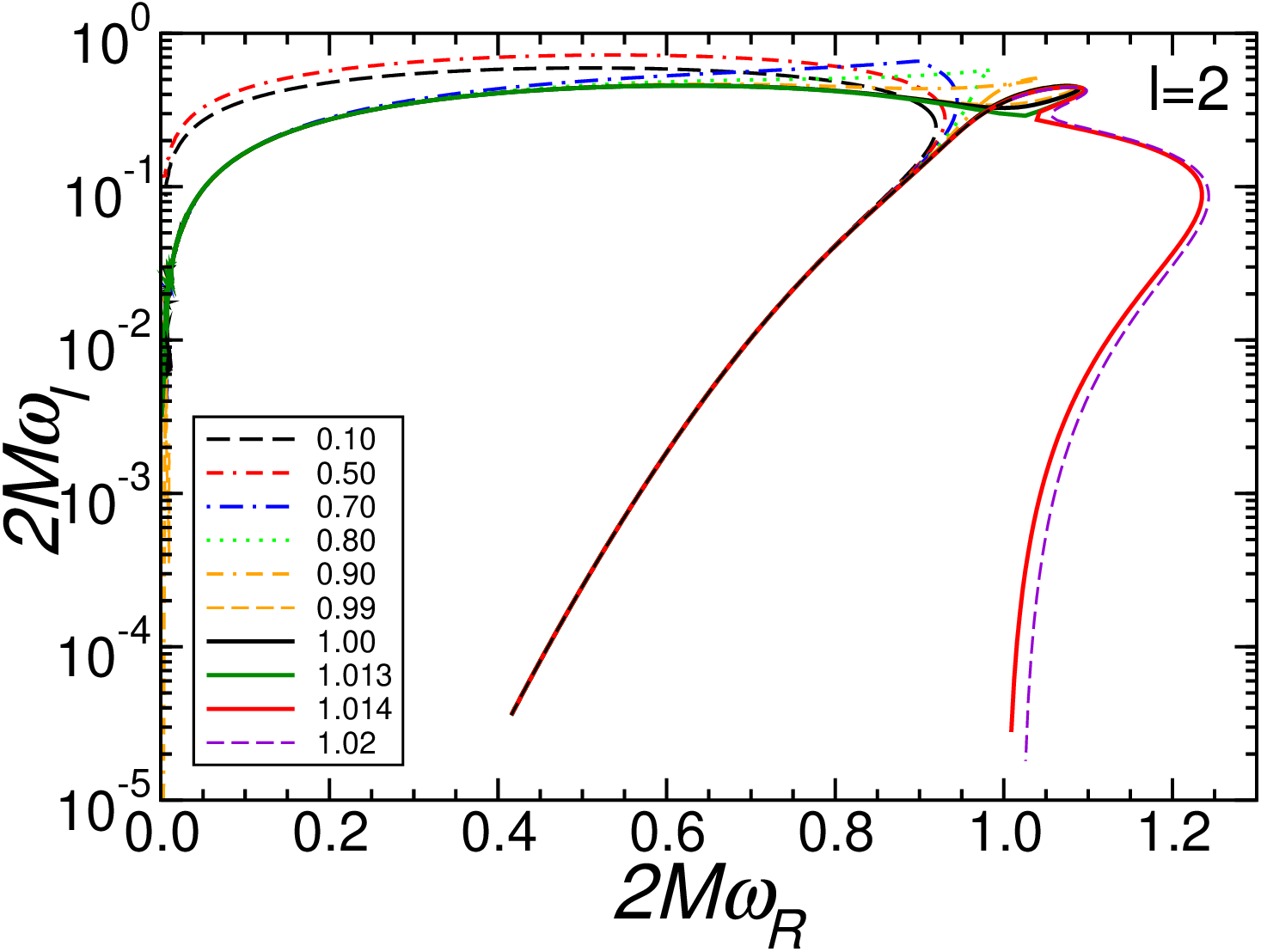}
\end{tabular}
\caption{Left: spectrum of the weakly damped family of QNMs. The vertical line
  corresponds to twice the orbital frequency of a particle in circular orbit
  at the ISCO: as we will discuss in a follow-up paper, only QNMs to the left
  of the line can be excited by a compact object inspiralling into the
  gravastar along quasi-circular orbits. In the case $v_s=0.8$ the mode
  ``turns around'' describing a loop in the complex plane. For $v_s\lesssim
  0.8$ the modes move {\it clockwise} in the complex plane as $\mu$
  increases. For $v_s\gtrsim 0.8$ they move {\it counterclockwise} and they
  cross the real axis at finite compactness. To facilitate comparison, in the
  right panel we show again the right panel of Fig.~\ref{fig:polarQNMsv} using
  a logarithmic scale for the imaginary part.
%  (compare the right panel of Fig.~\ref{fig:modesreim} below).
}
\label{fig:newmodes}
\end{figure*}
\end{center}

\begin{center}
\begin{figure*}[ht]
\begin{tabular}{cc}
\includegraphics[width=8.8cm,clip=true]{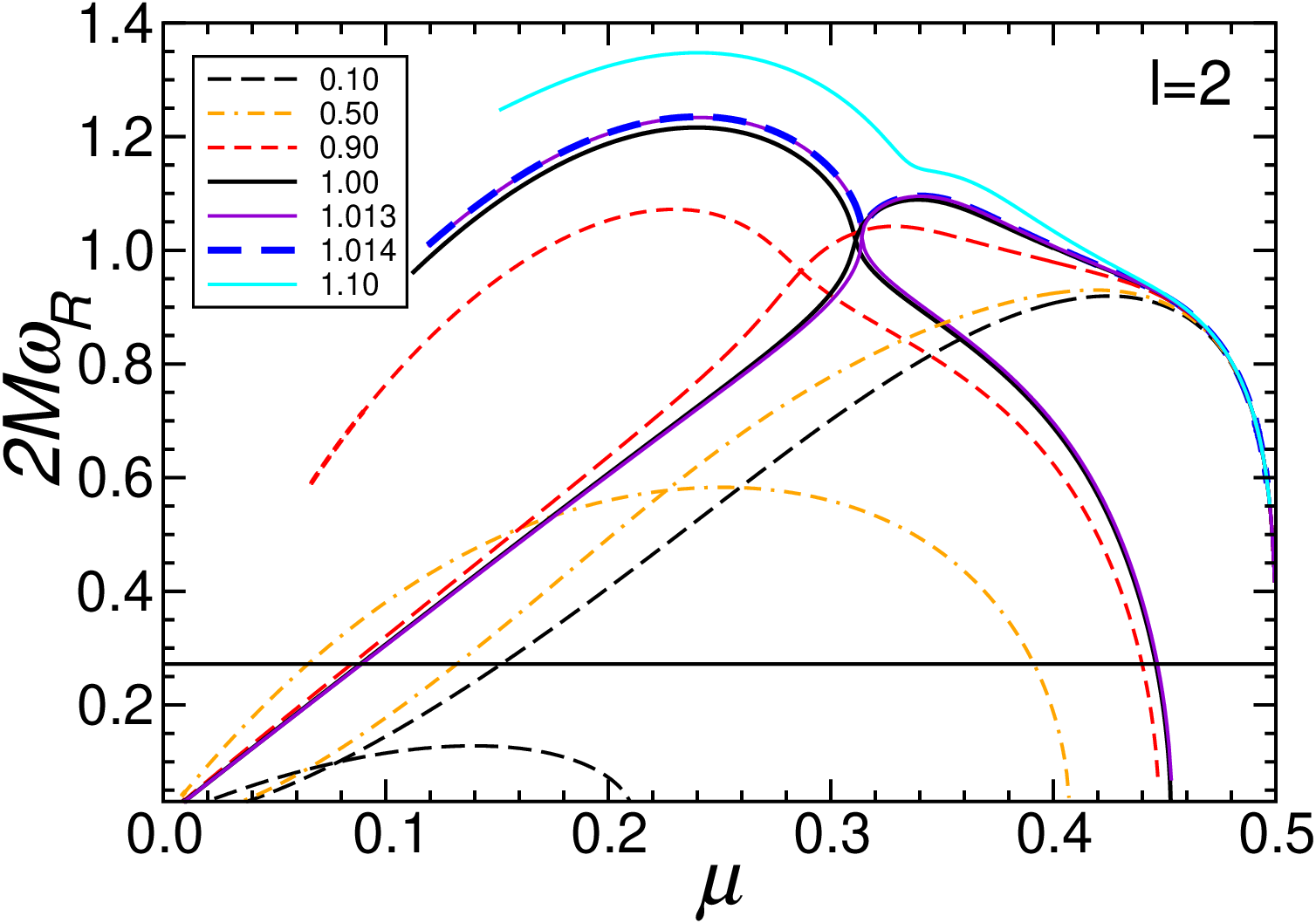}&
\includegraphics[width=8.8cm,clip=true]{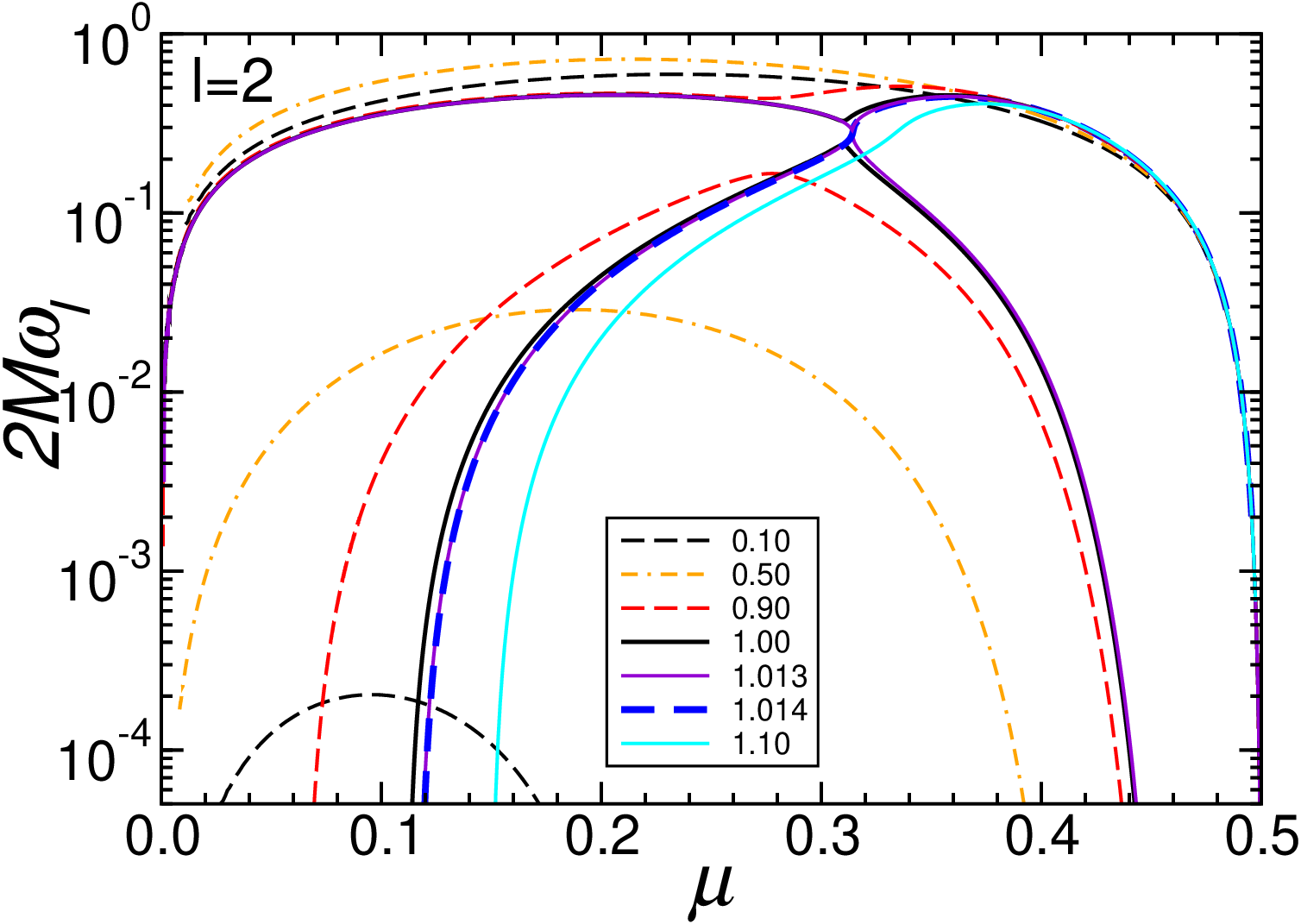}\\
\end{tabular}
\caption{Real (left) and imaginary parts (right) of the fundamental polar
  $w$-mode for different values of the equation of state parameter
  $v_s$. Different linestyles correspond to different values of $v_s^2$, as
  indicated in the legend. The horizontal line in the left panel corresponds
  to twice the orbital frequency of a particle in circular orbit at the ISCO:
  only modes below the line can be excited during a quasicircular inspiral.}
\label{fig:polarQNMsreim}
\end{figure*}
\end{center}

The $v_s$-dependence of the modes is studied in Figs.~\ref{fig:polarQNMsv},
\ref{fig:newmodes} and \ref{fig:polarQNMsreim}.  In Fig.~\ref{fig:polarQNMsv}
we show the tracks described in the complex plane by the fundamental polar and
axial $w$-mode as we vary the compactness parameter $\mu$.  The fundamental
axial mode does not depend on the equation of state parameter, as expected,
but the polar modes do change as a function of $v_s$. The standard argument
used to deduce that the speed of sound $v_s^2>0$ does not necessarily hold
when one deals with exotic matter (as in the case of gravastars and
wormholes). Therefore, for completeness, in the left panel of
Fig.~\ref{fig:polarQNMsv} we compute polar QNMs when $v_s^2<0$.  Different
linestyles correspond to different values of $v_s^2$, as indicated in the
legend. The solid black line reproduces the fundamental axial $w$-mode of
Figs.~\ref{fig:axpolQNMs} and \ref{fig:modes}. The dashed black line
corresponds to the fundamental polar $w$-mode for a shell with low sound speed
($v_s^2=0.1$), corresponding to the fundamental polar $w$-mode of
Figs.~\ref{fig:axpolQNMs} and \ref{fig:modes}. The dash-dash-dotted (red),
dash-dotted (blue) and dotted (green) lines represent a marginally subluminal,
imaginary sound speed ($v_s^2=-0.99$) and superluminal sound speeds
($v_s^2=-2$ and $v_s^2=-4$, respectively). Nothing particularly striking
happens in this regime: QNM frequencies for polar and axial perturbations are
different in all cases, but for large compactness the results become
$v_s$-independent and modes of different parity become approximately
isospectral, in agreement with the analytical results of Appendix
\ref{app:highC}.  Furthermore, as $|v_s|\to \infty$ the polar modes seem to
approach the axial mode. We can perhaps understand this behavior if we think
that the shell is effectively becoming so stiff that matter decouples from the
space-time dynamics, and only the ``space-time'' character of the oscillations
survives.

The situation is more interesting in the case $v_s^2>0$, displayed in the
right panel of Fig.~\ref{fig:polarQNMsv}. At first (when $v_s^2\leq 0.5$ or
so) the modes show a behavior similar to that seen for $v_s^2<0$, albeit in
the opposite direction (i.e. the real and imaginary parts of the QNM
frequencies {\it increase} rather than decreasing when $|v_s|$
increases). When $v_s^2=0.7$ a cusp develops, and as the speed of sound
approaches the speed of light (for $v_s^2\gtrsim 0.9$ in the figure) the modes
``turn around'' describing a loop in the complex plane. The area of this loop
in the complex plane increases until the sound speed reaches a critical value
$1.013\leq v_{\rm crit}^2\leq 1.014$ (corresponding to $v_{\rm crit}\simeq
1.007$). For $v_s>v_{\rm crit}$ the QNM behavior changes quite
drastically. The complex mode frequencies still approach the $\omega_I=0$ axis
clockwise as $\mu \to 0.5$. However, as $\mu$ decreases the modes approach the
axis $\omega_I=0$ very rapidly along tracks which are now tangent to the {\it
  lower} branch of the fundamental axial mode.

Even more interestingly, when $v_s^2>0$ there is also a {\em second} family of
QNMs with very small imaginary part. A plot of the eigenfunctions shows that
these modes are similar in nature to the $w_{\rm II}$-modes.  The trajectories
described in the complex plane by some of these weakly damped QNMs are shown
in the left panel of Fig.~\ref{fig:newmodes}. For comparison, the right panel
of Fig.~\ref{fig:newmodes} shows some of the ``ordinary'' modes. These
ordinary modes are the same as in the right panel of
Fig.~\ref{fig:polarQNMsv}, except that this time we use a logarithmic scale
for the imaginary part. The second family of QNMs plotted in the left panel
has very long damping, and in this sense it is similar to the $s$-modes of
``ordinary'' ultra-compact stars discussed by Chandrasekhar and Ferrari
\cite{ChandrasekharFerrari}. There is, however, an important difference:
unlike the $s$-modes, which exist only when a star is extremely compact, the
weakly damped modes of a thin-shell gravastar only exist for {\em small}
compactness $\mu<\mu_{\rm crit}$.

In the left (right) panel of Fig.~\ref{fig:polarQNMsreim} we show the real
(imaginary) parts of both families of QNMs as functions of $\mu$ for selected
values of $v_s$.  Both the real and imaginary part of the weakly damped modes
tend to zero at some finite, $v_s$-dependent compactness $\mu$. The range
where weakly damped modes exist increases with $v_s^2$. The fact that both the
mode frequency and its damping tend to zero at the critical compactness
$\mu_{\rm crit}$ suggests that the mode somehow ``disappears'' there, rather
than undergoing a nonradial instability, but this conjecture deserves a more
careful analytical study. Plots similar to those shown in
Fig.~\ref{fig:polarQNMsreim} show that the imaginary part of ``ordinary''
modes with $v_s^2\gtrsim 0.84$ ($v_s\gtrsim 0.92$) rapidly approaches zero at
some finite compactness $\mu$ while the real part of the modes stays finite.
We are unable to track QNMs numerically when $2M\omega_I\lesssim 10^{-5}$
using the continued fraction method,
%with the accuracy parameters used for this calculation,
and in any case we cannot really {\it prove} by numerical methods that
$\omega_I\to 0$ at some finite compactness $\mu>0$. When $v_s^2=1.0134$ the
two family of modes approach each other along a cusp, but their frequencies
and damping times never cross. The ``ordinary'' family of modes exists all the
way up to $\mu=0.5$ but it becomes unstable (in the sense that the imaginary
part of the mode crosses the real axis with the real part remaining finite) at
some finite, $v_s$-dependent value of the compactness $\mu$. 

Summarizing, our numerical results suggest that (1) weakly damped QNMs only
exist when their compactness is smaller than some $v_s$-dependent critical
threshold and (2) when $v_s>0$ and $v_s^2\gtrsim 0.84$, the imaginary part of
a QNM crosses the real axis at another critical threshold whose value can be
estimated by extrapolation. Nonrotating thin-shell gravastars should be {\it
  unstable} against nonradial perturbations when their compactness is smaller
than this critical value.

\begin{center}
\begin{figure}[ht]
\includegraphics[width=8.8cm,clip=true]{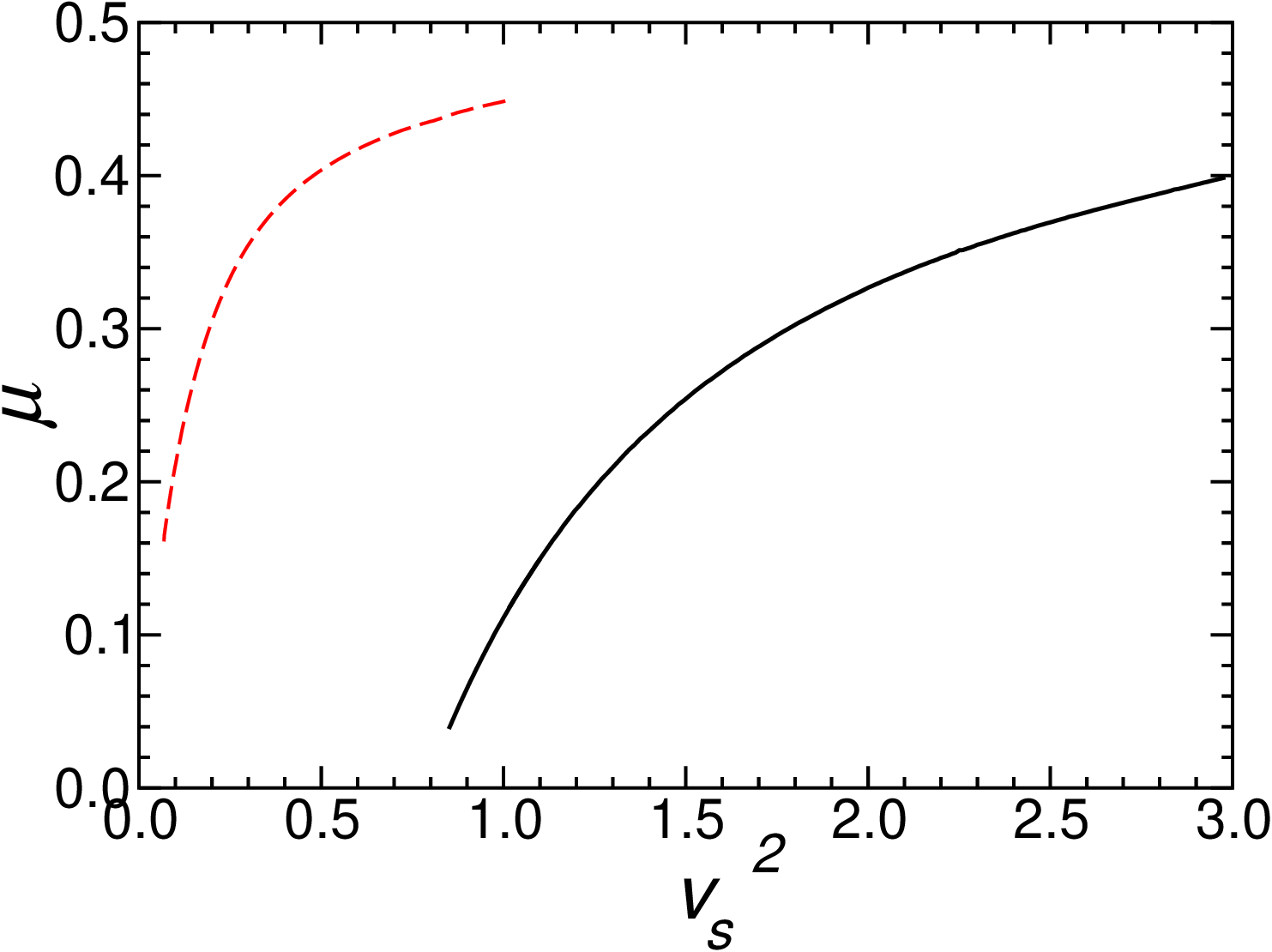}
\caption{Significant thresholds in the $(\mu,\,v_s^2)$ plane. Thin-shell
  gravastars should be unstable with respect to nonradial perturbations below
  the solid line, corresponding to ``ordinary'' QNM frequencies whose
  imaginary part crosses the zero axis while their real part stays finite. The
  dashed line corresponds to the ``vanishing point'' of weakly damped QNM
  frequencies, i.e. to the point where both their real and imaginary parts
  have a zero crossing. We could not find weakly damped modes in the region
  above the dashed line.}
%{\bf [E: here we should add the region of existence of the ``$s$-modes'' and
%    the instability line for $l=3$. If it turns out to be different and more
%    stringent, then maybe we will also need higher $l$'s? I still don't see
%    what is special with $v_s^2=0.84$.]}
\label{fig:vmucrit}
\end{figure}
\end{center}

The two thresholds are plotted in Fig.~\ref{fig:vmucrit}. Weakly damped QNMs
with $l=2$ exist only below the dashed line and,
%It is very hard to track the instability for $v_s\lesssim 0.85$, but we
%suspect that no instability should exist for $v\lesssim 0.84$ or so.
according to our extrapolation of numerical results, thin-shell gravastars
should be unstable to nonradial perturbations with $l=2$ below the solid line
(we verified that the instability condition for $l=3$ is less stringent than
for $l=2$). The dashed line extends up to $v_s^2< 1.0134$, where our numerical
search for weakly damped QNMs becomes impractical (the modes trace smaller and
smaller loops in the complex plane and they seem to disappear when the
compactness is still smaller than $\mu=0.5$).
%
%{\bf [E: can we say something about $l=3$ thresholds? They should be less
%    stringent.]}

To validate results from the continued fraction method we used an independent
numerical approach: the resonance method
\cite{ChandrasekharFerrari,Chandrasekhar:1992ey,Berti:2009wx}, which is
applicable to QNMs with $\omega_I\ll \omega_R$. The resonance method was first
used by Chandrasekhar and Ferrari in their analysis of gravitational wave
scattering by constant-density, ultra-compact stars
\cite{ChandrasekharFerrari,Chandrasekhar:1992ey}. Chandrasekhar and Ferrari
showed that the radial potential describing odd-parity perturbations of these
stars displays a local minimum as well as a maximum when the stellar
compactness $\mu\gtrsim0.39$. If this minimum is sufficiently deep,
quasi-stationary, ``trapped'' states can exist: gravitational waves can only
leak out to infinity by ``tunneling'' through the potential barrier. The
damping time of these modes is very long, so they were dubbed ``slowly
damped'' modes (or $s$-modes) \cite{ChandrasekharFerrari}.

\begin{center}
\begin{figure}[ht]
\begin{tabular}{c}
\includegraphics[width=8.8cm,clip=true]{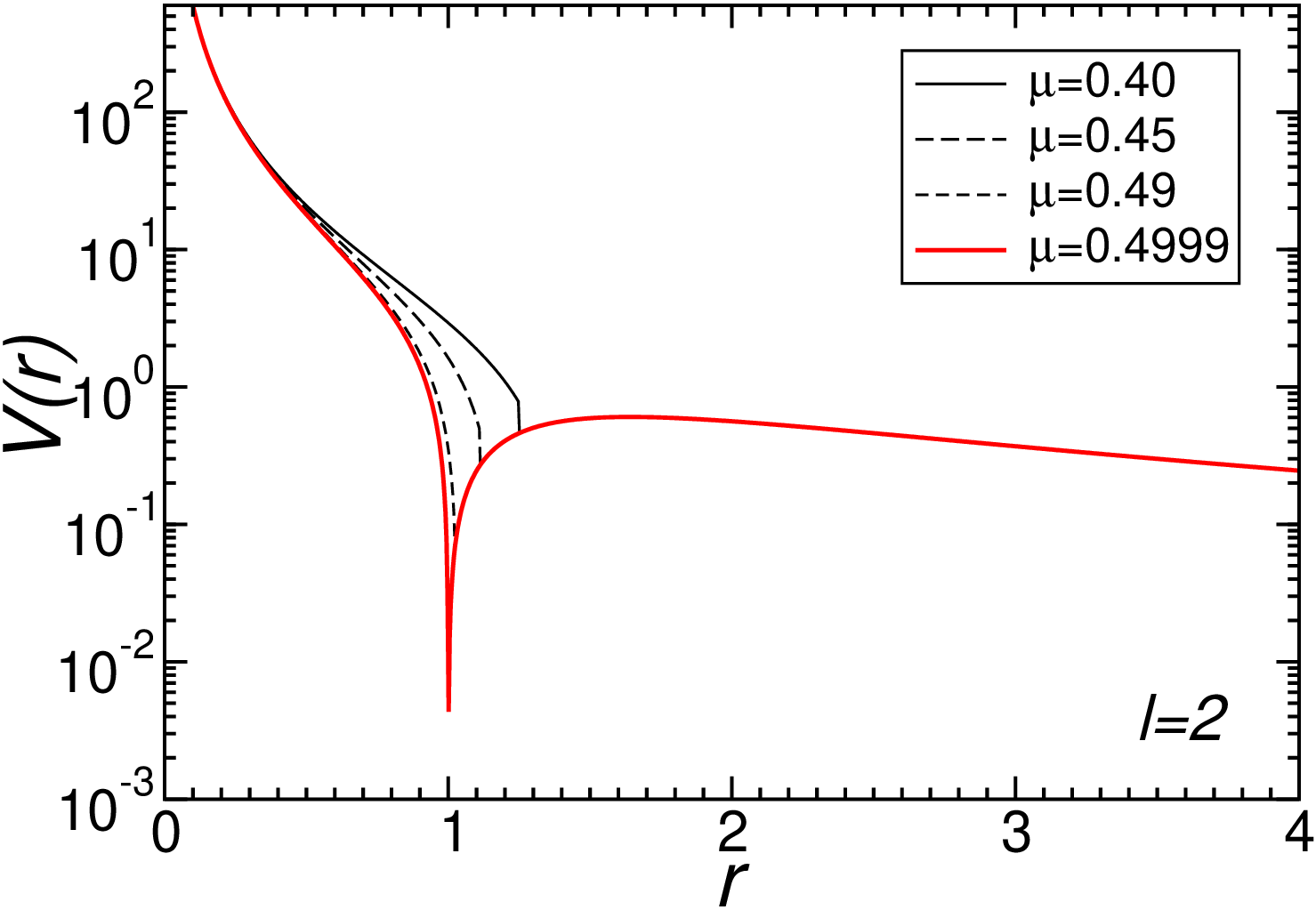}
\end{tabular}
\caption{The potential governing axial and polar perturbations for different
  values of the gravastar compactness $\mu\equiv M/a$, where $a$ is the
  location of the shell (see \cite{ChandrasekharFerrari}, showing a similar
  plot for axial perturbations of constant-density stars). The potential
  develops a minimum when $\mu\gtrsim 0.43$. Note that the polar and axial
  perturbations in the interior are both governed by the {\em same} potential,
  given in Eq.~\ref{eq:potin} below.}
\label{fig:plot}
\end{figure}
\end{center}

A straightforward analysis of Eq.~(\ref{eq:potin}) and an inspection of
Fig.~\ref{fig:plot} show that the axial potential for a gravastar develops a
minimum when $\mu\gtrsim0.43$. The compactness of ordinary stars is limited by
the Buchdal limit ($\mu<4/9\simeq 0.4444$), but since gravastars can be
considerably more compact than this limit, $s$-modes can exist all the way
down to the ``Schwarzschild limit'' ($\mu\to 0.5$).  These modes can be
computed via the continued fraction method, but since they are long-lived the
resonance method, which is computationally very simple, provides very accurate
estimates of their frequencies and damping times.  We find that the resonance
method and the continued fraction method are in very good agreement whenever
the resonance method is applicable.  In the limit $\mu\to 1/2$ we have ${\rm
  Im}(\omega)\ll {\rm Re}(\omega)$, and all QNM frequencies can be interpreted
as ``trapped'' states.
%{\bf [Y: According to Poisson and Sasaki \cite{Poisson:1994yf} (or the
%    perturbation theory chapter in Novikov and Frolov \cite{Frolov:1998wf}),
%    the transmissivity of the barrier is
%    $[(l!)3/[(2l)!(2l+1)!]]2(4\omega)^{2l+2}$. For $l=2$, and $\omega=0.13$,
%    this is $1.5\times 10^{-7}$, and this could be the ratio between the
%    imaginary and real parts.}

The resonance method essentially confirms our continued fraction results for
modes with $\omega_I\ll \omega_R$. In particular, it provides additional
numerical evidence for the conjectured nonradial instability of thin-shell
gravastars with low compactness. Despite the numerical evidence, an analytic
confirmation of our estimates of the instability threshold would be highly
desirable.

%

%%%%%%%%%%%%%%%%%%%%%%%%%%%%%%%%%%%%%%%%%%%%%%%%%%%%%%%%%%%%%%%%%%%%%%%%%%%%%%%%%%%%%%%%%%%%%%%%%%%%%%%%%%%%%%%%%%%%%%
\section{Conclusions and outlook}
%%%%%%%%%%%%%%%%%%%%%%%%%%%%%%%%%%%%%%%%%%%%%%%%%%%%%%%%%%%%%%%%%%%%%%%%%%%%%%%%%%%%%%%%%%%%%%%%%%%%%%%%%%%%%%%%%%%%%%

In this work we have studied the nonradial perturbations on nonrotating,
thin-shell gravastars. It should not be too hard to extend our formalism to
the more complex case of five-layer gravastars of the type originally proposed
by Mazur and Mottola (see \cite{Chirenti:2007mk} for a treatment of the axial
case).

A presumably less trivial extension concerns rotating gravastars.  Slowly
rotating gravastars may be unstable against scalar perturbations because of an
exponential growth of the perturbations due to superradiance, the so-called
``ergoregion instability'' \cite{Cardoso:2007az,Chirenti:2008pf}. An extension
of the present formalism can be used to study nonradial gravitational
perturbations of slowly rotating gravastars and to discuss their ergoregion
instability, which is believed to be stronger for gravitational perturbations
\cite{Cardoso:2007az}. For gravastars this instability is due to superradiant
gravitational wave scattering in the ergoregion, so it is essentially the same
as the ``$w$-mode instability'' discussed by Kokkotas {\it et al.}  for
ultracompact stars \cite{Kokkotas:2002sf}. The main difference is that
gravastars can be more compact than constant-density stars, so we may expect
the instability to be stronger.

Finally, our formalism can be applied to explore nonradial gravitational
perturbations of nonrotating wormholes, where the position of the throat plays
a role similar to the thin shell of a gravastar. Such an analysis could
confirm or disprove some conjectures on the similarity of the QNM spectra of
wormholes and BHs \cite{Damour:2007ap}.

In follow-up work we will extend our study to explore QNM excitation by
compact objects in closed orbits around thin-shell gravastars (see
\cite{Andrade:1999mj,Ferrari:2000sr} for similar studies considering the
excitation of axial modes by scattering orbits). According to preliminary
estimates by Norte, following the weak-gravity expansion of Poisson and Sasaki
\cite{Poisson:1994yf} (extended by Li and Lovelace \cite{Li:2007qu} to general
boundary conditions), the gravitational wave luminosity can change
dramatically at the resonances, while being very close to the BH value in
nonresonant conditions. A weak-field study should provide a reasonably
accurate estimate of the orbital frequency at which the resonance occurs, but
it can only provide a leading-order estimate of the radiated energy.  A more
precise characterization requires the numerical integration of the
perturbation equations
\cite{Kojima:1987tk,Gualtieri:2001cm,Pons:2001xs,Berti:2002ry}. In a follow-up
paper we will compare gravitational radiation from EMRIs in thin-shell
gravastar space-times to the BH results of
Refs.~\cite{Tanaka:1993pu,Poisson:1993vp,Cutler:1993vq,Cutler:1994pb}.

%%%%%%%%%%%%%%%%%%%%%%%%%%%%%%%%%%%%%%%%%%%%%%%%%%%%%%%%%%%%%%%%%%%%%%%%%%%%%%
\section*{Acknowledgements}
%%%%%%%%%%%%%%%%%%%%%%%%%%%%%%%%%%%%%%%%%%%%%%%%%%%%%%%%%%%%%%%%%%%%%%%%%%%%%%
This work was partially supported by FCT - Portugal through projects
PTDC/FIS/64175/2006, PTDC/FIS/098025/2008 and PTDC/FIS/098032/2008. Y.C. was
supported by NSF grants PHY-0653653 and PHY-0601459, and the David and Barbara
Groce Start-up Fund at Caltech. E.B.'s research was supported by NSF grant
PHY-0900735.
%%%%%%%%%%%%%%%%%%%%%%%%%%%%%%%%%%
%%%%%%%%%%%%%%%%%%%%%%%%%%%%%%%%%%
\appendix

%%%%%%%%%%%%%%%%%%%%%%%%%%%%%%%%%%%%%%%%%%%%%%%%%%%%%%%%%%%%%%%%%%%%%%%%%%%%%%%%%%%%%%%%%%%%%%%%%%%%%%%%%%
\section{\label{app:match}Perturbation equations and matching conditions}
%%%%%%%%%%%%%%%%%%%%%%%%%%%%%%%%%%%%%%%%%%%%%%%%%%%%%%%%%%%%%%%%%%%%%%%%%%%%%%%%%%%%%%%%%%%%%%%%%%%%%%%%%%
In this appendix we develop the formalism to study polar and axial nonradial
(linear) oscillations of an object consisting of a thin spherical shell
separating two spherically symmetric regions.  Though we are mainly interested
in thin-shell gravastars, we shall keep the discussion as general as
possible. We focus on a background metric of the form (\ref{eq:g0}), keeping
$f(r)$ and $h(r)$ generic.

The Regge-Wheeler gauge is incompatible with the requirement that the shell's
world tube sits at a fixed location. We deal with this issue in two steps:
(i) we choose a coordinate system (system A) such that perturbations in both
the interior and exterior are in the Regge-Wheeler gauge;
(ii) we write down the equations of motion of mass elements on the shell in
this coordinate system (separately for the interior and the exterior), and try
to apply matching conditions on the boundary of this moving shell.  We then
transform to a new coordinate system (system B) in which the shell is fixed,
simplifying the matching procedure. System B is an auxiliary tool for
matching: when we consider perturbations of the gravastar induced (say) by
orbiting particles we will mostly use system A, which is the usual
Regge-Wheeler gauge.

Step (i) is a straightforward adaptation of formulas in Section
\ref{sec:pert}. In the Regge-Wheeler gauge, polar perturbations are defined by
three functions $K$, $H$ and $H_1$ for the interior and exterior,
respectively.  These functions satisfy an algebraic relation and two coupled
ODEs, which can be used to reduce the problem to a wave equation. This means
that we only need two conditions relating these quantities from the two
sides. Similarly, for axial perturbations we have $h_0$ and $h_1$, which
satisfy a coupled wave equation, and again we need only two conditions
connecting the interior and exterior.

Now we go directly to step (ii), and parametrize the shell position as in
Eqs.~(\ref{shellpos}).  Let us define
\beq
\hat{t} &\equiv& \tau/\sqrt{f(a)}\,,\\
\dot{F} &\equiv &( \partial F/\partial\hat t) = 
\sqrt{f(a)} ({\partial F}/{\partial \tau})\,.
\eeq
Note that $\hat{t}$ is simply a re-scaling of proper time of the mass element,
and that $f(a)$ is common to the interior and exterior, due to the requirement
that the induced metric is continuous.  In the absence of perturbations, $\hat
t$ coincides with $t$. Henceforth, we will use $(\hat t,\theta_*,\varphi_*)$,
to parametrize the mass element.  As a consequence, the four-velocity
$u^\alpha$ of the mass element $(\theta_*,\varphi_*)$ would be as in
Eq.~(\ref{4vel}).
Imposing that $g_{\alpha\beta} u^{\alpha}u^{\beta}=-1$, we actually have to
require that
\be
\frac{ (1+2\delta\dot{t})g_{tt}(\hat t + \delta t, a+\delta r,\theta_*+
\delta\theta ,\varphi_*+\delta\varphi)  }{f(a)}= -1\,,
\ee
which, to leading order, is
\beq
&&f'(a) \delta r(\hat t,\theta_*,\varphi_*)+ 2 f(a)
\delta\dot  t(\hat t,\theta_*,\varphi_*) \nonumber\\
&-&\delta_{\rm RW} g_{tt}(\hat t,\theta_*,\varphi_*)
=0\,.
\eeq
Here $\delta_{\rm RW} g_{tt}$ is the $tt$-component of the metric perturbation
in the Regge-Wheeler gauge.

We will now carry out a gauge transformation, both in the exterior and in the
interior, which maps the shell to a fixed sphere. As explained in the main
text, for any general gauge transformation $\bar x^{\bar\alpha} =
x^\alpha-\xi^{\alpha}(x^\mu)$ we have
\beq
g_{\alpha\beta}(x^\mu) &=& \bar{g}_{\mu\nu}(\bar x^\rho) 
\left(\frac{\partial \bar x^\mu}{\partial  x^\alpha}\right)
\left(\frac{\partial \bar x^\nu}{\partial  x^\beta}\right)
\nonumber\\
&=&
\left[\bar{g}_{\alpha\beta} 
-\bar{g}_{\alpha\nu} {\xi^\nu}_{,\beta}
-\bar{g}_{\mu\beta} {\xi^\mu}_{,\alpha}\right]_{\bar x^\rho},
\eeq
where we have ignored terms of second order in $\xi^\mu$. Noting that 
\begin{equation}
g_{\alpha\beta}(x^\mu) =g_{\alpha\beta}(\bar x^\mu + \xi^\mu)\,,
\end{equation}
we obtain Eq.~(\ref{deltagmunu}). This is the desired form, because we want to
use $\xi^\mu$ to transform to a coordinate system $\bar x^\mu$ where mass
elements on the shell are fixed in spatial location and move uniformly in the
time direction: i.e., for mass elements on the shell, $\bar x^\mu = (\hat
t,a,\theta_*,\varphi_*)$. Then Israel's junction conditions will be applied to
the Regge-Wheeler metric evaluated at a fixed coordinate location $(\hat
t,a,\theta_*,\varphi_*)$, and in terms of the transformation generators (which
are related to the motion of the shell in the Regge-Wheeler gauge).

Let us make four consecutive transformations
\begin{eqnarray}
\xi_{(0)}^\alpha &=& [-f^{-1}(r) y(t)  Y_{lm},0,0,0]\,,\nonumber\\ 
\xi_{(1)}^\alpha &=&[0, h(r) z(t)  Y_{lm},0,0] \,, \nonumber\\
 \xi_{(2)}^\alpha &=&\left[0, 0, \frac{w(t)}{r^2} Y_{lm,\theta} ,
\frac{w(t)}{r^2\sin^2\theta} Y_{lm,\varphi}\right] \,, \nonumber\\
 \xi_{(3)}^\alpha&=&\left[0, 0, \frac{x(t)}{r^2 \sin\theta} Y_{lm,\varphi} ,
-\frac{x(t)}{r^2\sin\theta} Y_{lm,\theta}\right]\,,\nonumber
 \end{eqnarray}
or, by lowering indices,
\begin{eqnarray}
\xi_{(0) \alpha} &=&[y(t)  Y_{lm},0,0,0] \,,\nonumber\\
\xi_{(1) \alpha} &=&[0,z(t)  Y_{lm},0,0] \,, \nonumber\\
 \xi_{(2) \alpha} &=&\left[0, 0, w(t) Y_{lm|\theta} ,w(t)  
Y_{lm|\varphi}\right] \,, \nonumber\\
 \xi_{(3) \alpha} &=&\left[0,0,  x(t) \epsilon_{\theta \varphi} {Y_{lm}}^{|\varphi},
 x(t) \epsilon_{\varphi \theta} {Y_{lm}}^{|\theta}\right]\,.\nonumber
 \end{eqnarray}
where the covariant derivative $|$ is defined with respect to the
2--dimensional metric
\begin{equation}\label{2dmetric}
G_{ab} = d\theta^2 + \sin^2\theta d\varphi^2\,.
\end{equation}
We impose that, when evaluated at $(t,r,\theta,\varphi) = (\hat t,
a,\theta_*,\varphi_*)$, the vector $\xi^\alpha$ coincides with $(\delta
t,\delta r,\delta\theta,\delta\varphi)$, so that in the new coordinate system
Eqs.~(\ref{shellpositionbar}) will be valid.
Such a transformation leads to the following changes in the metric components:
\begin{equation}
\delta_{(0)} g_{\mu\nu}  =  
\left[
\begin{array}{cc|cc}
 2 \dot{y}  & - \frac{f' }{f} y& y \partial_{\theta} & y \partial_{\varphi}   \\
  * &   &  \\
 \hline
* & & \\
*& &  \\
\end{array}
\right]Y_{lm}\,,
\end{equation}
\begin{equation}
\delta_{(1)} g_{\mu\nu}  =  
\left[
\begin{array}{cc|cc}
- f' h z &  \dot{z} &  \\
 * & \frac{h'}{h} z   &  z \partial_{\theta}   &  z \partial_{\varphi}  \\
 \hline
&  * & 2 r h  z & \\
&  * &  &  2 r h   z\sin^2\theta 
\end{array}
\right]Y_{lm}\,,
\end{equation}
which are purely polar perturbations;
\begin{equation}
\delta_{(2)} g_{\mu\nu}  =  
\left[
\begin{array}{cc|cc}
& & \dot{w} Y_{lm,\theta} & \dot{w} Y_{lm,\varphi}\\
& & - \frac{2w}{r} Y_{lm,\theta} & -\frac{2w}{r} Y_{lm,\varphi}\\ \hline
*&*  & 2 w Y_{lm|\theta\theta} & 2 w  Y_{lm|\theta\varphi} \\
*&* &  * &   2 w Y_{lm|\varphi\varphi}
\end{array}
\right]\,,
\end{equation}
which are also polar perturbations, and finally
\begin{equation}
\delta_{(3)} g_{\mu\nu}  =  
\left[
\begin{array}{cc|cc}
& & 
\dot{x}\epsilon_{\theta\varphi}{Y_{lm}}^{|\varphi} &
\dot{x}\epsilon_{\varphi\theta}{Y_{lm}}^{|\theta}\\
& & 
-  \frac{2x}{r} \epsilon_{\theta\varphi}{Y_{lm}}^{|\varphi} &
- \frac{2x}{r} \epsilon_{\varphi\theta}{Y_{lm}}^{|\theta}\\
\hline
* & 
* & 
x \phi_{\theta\theta} & x \phi_{\theta\varphi} \\
*& 
* & 
x \phi_{\varphi\theta} & x \phi_{\varphi\varphi}
\end{array}
\right]\,,
\end{equation}
which are axial perturbations.  Here we have defined
\begin{equation}
\phi_{mn} = 
{\epsilon_{m}}^{a} Y_{lm|na}+
{\epsilon_{n}}^{a} Y_{lm|ma}\,,
\end{equation}
where $m$, $n$ and $a$ run through $\theta$ and $\varphi$.  Here
$\epsilon_{ab}$ is again defined with respect to the metric (\ref{2dmetric}),
so that
\begin{equation}
\epsilon_{\theta\varphi}= -\epsilon_{\varphi\theta} =\sin\theta\,.
\end{equation}

In terms of $y$ and $z$, the normalization of the 4--velocity would be written
as
\beq
0&=& -f'(a) h(a) z(\hat t) + f(a) H(\hat t,a) -  
2 f(a) [-f^{-1}(a) \dot{y}(\hat t)]\nonumber \\
&=&-f'(a) h(a) z(\hat t) + f(a) H(\hat t,a) + 2 \dot{y}(\hat t)\,.\nonumber
\eeq
Here we note again that $f_+=f_-$. In this new coordinate system the metric is
given by Eq.~(\ref{gbarmunu}), where
\begin{equation}
\delta g _{\alpha\beta}= \delta_{(1)} g _{\alpha\beta}+ 
\delta_{(2)} g _{\alpha\beta}+ \delta_{(3)} g _{\alpha\beta}\,,\nonumber
\end{equation}   
We are now in a position to match metric components along the shell, which
simply sits at $(\hat t,a,\theta_*,\varphi_*)$.  Of course, all of the
matching conditions will be expressed in terms of the Regge-Wheeler metric
perturbations, and the motion of the shell in the (internal and external)
Regge-Wheeler gauges. In the $\theta$ and $\varphi$ directions, we have
\beq
x_+ (\hat t)&=& x_-(\hat t),\\
 w_+(\hat t)  & =& w_-(\hat t)\,,\\
 2 h_+(a) z_+(\hat t)  + a K_+(\hat t,a) 
& =& 2  h_- (\hat t)  z_- (\hat t) + a K_-(\hat t,a)\,.\nonumber\\
\eeq
In the $t\theta$ and $t\varphi$ directions we have, in addition,
\begin{equation}
y_+(\hat t) = y_-(\hat t)\,,\quad h_{0+}(\hat t,a) = h_{0-}(\hat t,a)\,.
\end{equation}
while in $tt$ direction the matching condition is automatically satisfied,
with
\begin{equation}
\bar g_{tt}(\hat t,a,\theta_*,\varphi_*) =-f(a)\,,
\end{equation}
accurate up to first order in the perturbations.  This is a consistency check,
since we have imposed that the four-velocity of mass elements on the shell is
\begin{equation}\label{4vel0}
u^\alpha = (1/\sqrt{f(a)},0, 0,0)\,,
\end{equation}
which should have a norm of $-1$.  In simplified form, we have
\beq
&&[[x]]=[[w]]=[[y]]=[[h_0]]=[[2 h z + aK]]\nonumber\\
&=&[[2\dot{y}+fH-f'hz]]=0\,.
\eeq
The symbol ``[[ ...]]'', as defined by Eq.~(\ref{eq:defjump}), gives the jump
of any given quantity across the shell.

The four-velocity of mass elements on the shell is given by Eq.~(\ref{4vel0}).
The surface stress-energy tensor of the shell is
\beq
S_{jk}  &=& [\Sigma -\Theta+(\delta \Sigma - \delta \Theta)Y_{lm} ] 
u_j u_k \nonumber \\
&-&[\Theta+\delta\Theta\, Y_{lm}] \gamma_{jk}\,,
\eeq
where $j$ and $k$ go through $t$, $\theta$ and $\varphi$.  
Now let us try to evaluate the extrinsic curvature of the shell at the
location $(\hat t,a,\theta_*,\varphi_*)$.  First of all, we note that
\begin{equation}
n_\alpha = (0,1,0,0)/\sqrt{\bar g^{rr}}\,,\nonumber
\end{equation}
and the extrinsic curvature is 
\begin{equation}
K_{ij} = - \nabla_i n_j =- n_{j,i} +{\Gamma^{\mu}}_{ji} n_{\mu}\,,\nonumber
\end{equation}
where $i,j$ run through $t\,,\theta\,,\varphi$.  Since $n_\alpha$ only has a
nonzero $r$-component the first term vanishes, and we obtain
\begin{equation}
K_{ij} =\frac{1}{\sqrt{\bar g^{rr}}}{ \Gamma^{r}}_{ij}\,.\nonumber
\end{equation}

{}From the static configuration of the star, it is easy to obtain that
\beq
\|S_{jk}\| &=&
\left[\begin{array}{ccc}
\Sigma f(a) \\
& -a^2\Theta \\
& & -a^2\Theta\sin^2\theta
\end{array}
\right]\,,\nonumber\\
\|\bar K_{jk}\| &=& 
a \sqrt{h(a)} \left[\begin{array}{ccc}
-\frac{2f}{a^2} \\
& \left[1+\frac{af'}{2f}\right] \\
& &\left[1+\frac{af'}{2f}\right] \sin^2\theta
\end{array}
\right]\,,\nonumber
\eeq
where $j\,,k = t\,,\theta\,,\varphi$. From this we have relations
(\ref{eq:SigmaTheta}) for static gravastars.

Now let us focus on first-order quantities. The tensors $S_{jk}$ and $K_{jk}$
can each be decomposed into six terms: for example,
\begin{eqnarray}
S_{jk} &=& 
\left[\begin{array}{ccc}
S^{(1)}  \\
& S^{(2)}   \\
& &  S^{(2)} \sin^2\theta
\end{array}
\right]Y_{lm} \nonumber\\
&+&
S^{(3)} 
\left[\begin{array}{ccc}
 & \partial_{\theta}  & 
\partial_\varphi   \\
 \partial_\theta  \\
\partial_\varphi 
\end{array}\right]Y_{lm}  \nonumber \\
&+&
S^{(4)} 
\left[\begin{array}{ccc}
 &  {\csc\theta}\partial_{\varphi} & 
- \sin\theta\partial_\theta \\
{\csc\theta}\partial_\varphi \\
- \sin\theta\partial_\theta\end{array}\right]Y_{lm}  \nonumber\\
&+&  S^{(5)} 
\left[\begin{array}{ccc}
0 \\
& Y_{|\theta\theta} &  Y_{|\theta\varphi}\\
 & Y_{|\varphi\theta} &  Y_{|\varphi\varphi}
\end{array}\right]\nonumber\\
&+&
S^{(6)} 
\left[\begin{array}{ccc}
0 \\
& \phi_{\theta\theta} &  \phi_{\theta\varphi}\\
 & \phi_{\varphi\theta} &  \phi_{\varphi\varphi}
\end{array}\right]\,.
\end{eqnarray}
We have
\begin{eqnarray}
S^{(1)} &=& f \delta\Sigma  +[ (\Theta-2\Sigma) (f H - h z f' +2 \dot{y})]\,, \\
S^{(2)} &=& - a (a K \Theta + 2 h z\Theta  +a \delta \Theta)\,, \\
S^{(3)} &=& -\Sigma (y+\dot{w})\,,\\ 
S^{(4)} &=& \Sigma(h_0 -\dot{x})\,,\\
S^{(5)}&=&-2\Theta w\,,\\
S^{(6)}&=&-\Theta x\,,
\end{eqnarray}
and
\begin{eqnarray}
\bar K^{(1)} &=&\frac{f\sqrt{h}}{a}(H-a K') + 
f\sqrt{h}z\frac{\lambda_L+2h-a h'}{a^2},  \quad\quad\\
\bar K^{(2)}&=&\frac{a^2\sqrt{h}}{2}
\bigg\{K'-H'+\frac{2(\dot{H}_1+\ddot{z})}{f} \nonumber \\
&&\quad\quad\;\;+\left(1+\frac{a f'}{2f}\right)\frac{2K-H }{a}+\nn\\
&&\quad\quad\;\; +\bigg[\frac{2(h-\lambda_L)}{a^2}
+\frac{2f'h+fh'}{af}
\nonumber \\
&&\quad\quad\quad\;\;  +\frac{f f' h'-2 (f')^2 h+2 h f f''}{2f^2}
\bigg]z
\bigg\}, \\
\bar K^{(3)} &=&
\frac{\sqrt{h}}{a}\left[\frac{a H_1}{2}+{2y}+a\dot{z}+
\left(1+\frac{af'}{2f}\right)\dot{w}\right],\\
\bar K^{(4)} &=&\sqrt{h} \left[\frac{\dot x-2h_0}{a}
\left(1+\frac{af'}{2f}\right)+\frac{h_0'-\dot{h}_1}{2}\right],\\
%\end{eqnarray}
%
%\begin{eqnarray}
\bar K^{(5)} &=&\frac{2\sqrt{h}}{a}\left[\frac{a z}{2} +
w\left(1+\frac{af'}{2f}\right)\right],\\
\bar K^{(6)}&=& \sqrt{h} \left[-\frac{h_1}{2}+\frac{x}{a}
\left(1+\frac{af'}{2f}\right)\right]\,,
\end{eqnarray}
where $\lambda_L = -l(l+1)$.  Here $\bar K^{(4)}$ and $\bar K^{(6)}$ are axial
quantities. The junction condition on $\bar K^{(6)}$ yield
\begin{equation}
\LL\sqrt{h} h_1\RR =0,
\end{equation}
which together with $\LL h_0 \RR=0$ completes the junction conditions for
axial perturbations (the junction condition on $\bar K^{(4)}$ yields an
equation of motion for the variable $x$). For polar quantities, from $[[\bar
    K^{(5)}]]=8\pi S^{(5)}$ we have
\begin{equation}
[[\sqrt{h}z]]=0\,.
\end{equation}
Then matching $[[\bar K^{(3)}]]=8\pi S^{(3)}$ gives us an equation of
motion for $w$, while matching $[[\bar K^{(1,2)}]]=8\pi S^{(1,2)}$ yields
\begin{eqnarray}
&& \bigg[\bigg[\sqrt{h}\left(\frac{H}{a}- K'\right)\bigg]\bigg] 
+ \bigg[\bigg[\frac{2h}{a^2}-\frac{h'}{a}\bigg]\bigg]\sqrt{h}z \nonumber \\ 
&=& 8\pi \delta\Sigma\,, \\
&&\bigg[\bigg[\sqrt{h}\left(K'-H'+\frac{2\dot{H}_1}{f}\right)\bigg]\bigg] 
\nonumber \\
&-&\bigg[\bigg[\sqrt{h}\left(1+\frac{af'}{2f}\right)\frac{H}{a}\bigg]\bigg]
\nonumber \\
&+&\bigg[\bigg[\frac{h'}{a}-\frac{2h}{a^2}+\frac{f''h}{f}-
\frac{f'h'}{2f}\bigg]\bigg]\sqrt{h}z =-16\pi \delta\Theta\,.\quad\quad
\end{eqnarray}
The remaining equations are
\begin{eqnarray}
[[ K ]]&=&- 2 [[\sqrt{h}]]\sqrt{h} z/a =8\pi\Sigma  \sqrt{h} z\,,\\
{}[[H]]&=&\bigg[\bigg[\frac{f' \sqrt{h}}{f}\bigg]\bigg] \sqrt{h}z = 
8\pi (\Sigma-2\Theta)\sqrt{h} z\,, \\
\delta\Theta & = & -v^2 \delta \Sigma\,,
\label{eq:eos1}
\end{eqnarray}
where $v$ is defined as in Eq.~(\ref{eq:v}).

The formalism described above is more general than we need for a static
thin-shell gravastar. It can easily be adapted to more general horizonless
space-times and to static wormholes. For the Mazur-Mottolla gravastar, we
have:
\beq
&&\Sigma=0\,, \quad \Theta=-\frac{[[f']]}{16\pi\sqrt{f(a)}}\,,\nonumber\\
&&\LL f\RR=0=[[f'']]\,,\quad[[f']]=\frac{6M}{a^2}\,,\quad[[{f'}^2]]=
-\frac{12M^2}{a^4}\,.\nonumber
\label{jumpf}
\eeq
Using the equations of Sec.~\ref{sec:in} above together with these junction
conditions we obtain continuity conditions for the shell position,
$[[x]]=[[y]]=[[w]]=[[z]]=0$, and the matching conditions for the axial and
polar perturbations functions presented in the main text
[Eqs.~(\ref{junctionaxial}), (\ref{junctionpolar1}), (\ref{junctionpolar2})
  and (\ref{junctionpolar3})].

%%%%%%%%%%%%%%%%%%%%%%%%%%%%%%%%%%%%%%%%%%%%%%%%%%%%%%%%%%%%%%%%%%%%%%%%%%%%%
\section{\label{app:cf}The continued fraction method}
%%%%%%%%%%%%%%%%%%%%%%%%%%%%%%%%%%%%%%%%%%%%%%%%%%%%%%%%%%%%%%%%%%%%%%%%%%%%%

Our numerical search for the QNMs of gravastars is based on the continued
fraction method, as modified in \cite{Leins:1993zz,Benhar:1998au}.  The QNMs of
an oscillating gravastar are solutions of Eqs.~(\ref{eq:master}) and
(\ref{masterSch}) satisfying the boundary conditions imposed by physical
requirements: $\Psi$ should be regular at the origin, have the behavior of a
purely outgoing wave at infinity and satisfy the junction conditions discussed
in Section \ref{sec:match}.  The QNM frequencies are the (complex) frequencies
$\omega=\omega_R+i\omega_R$ for which these requirements are satisfied.

The numerical determination of the QNM frequencies is nontrivial, especially
for modes with large imaginary parts (strongly damped modes).  The reason is
simple to understand. Solutions of Eq.~(\ref{masterSch}) representing outgoing
and ingoing waves at infinity have the asymptotic behavior $\Psi^{\rm out}\sim
e^{r_*/\tau}$ and $\Psi^{\rm in} \sim e^{-r_*/\tau}$ as $r_*\to \infty$, where
$\tau=1/\omega_I$ is the damping time.  Therefore, identifying by numerical
integration the purely outgoing solutions (that is, those solutions for which
$\Psi^{\rm in}$ is zero) becomes increasingly difficult as the damping of the
mode increases.
The same problem occurs also in the case of QNMs of BHs, and was solved by
Leaver \cite{Leaver:1985ax}.  Leaver found a continued fraction relation that
can be regarded as an implicit equation which identifies the quasinormal
frequencies, thus circumventing the need to perform an integration out to
large values of $r_*$.  This method was subsequently adapted to the polar and
axial oscillations of a star \cite{Leins:1993zz,Benhar:1998au}.  The
Regge-Wheeler equation, which describes the perturbed space-time outside the
gravastar, becomes
\be\label{mastereq}
\frac{d^2 \Psi}{dr_*^2}+\left[\omega^2-  V_{\rm out}\right]\Psi=0\,,
\ee
with
\begin{equation}
V_{\rm out}=\left(1-\frac{2M}{r}\right)
\left(\frac{l(l+1)}{r^2}-\frac{6M}{r^3}\right)
\end{equation}
and the tortoise coordinate $r_*=r+2M\ln(r/2M-1)$.
We shall now write the solution of the Regge-Wheeler equation in a
power-series form as follows.  
Defining $z\equiv 1-R_2/r$, where $r=R_2$ is some point outside the shell of
the gravastar, and introducing a function $\phi(z)$, related to $\Psi(r)$ by:
\be\label{subst}
\Psi(r)=(r-2M)^{-i2M\omega}e^{-i\omega r}\phi(z)\equiv \chi(r)\phi(z)\,,
\ee
one finds that $\phi$ satisfies the differential equation:
\beq\label{stellLeav}
&&(c_0+c_1z+c_2 z^2+c_3 z^3)\frac{d^2\phi}{dz^2}+
(d_0+d_1 z+d_2 z^2)\frac{d\phi}{dz} \nonumber\\
&+&(e_0+e_1 z)\phi=0\,.
\eeq
The constants depend only on $\omega,\,l$ and $R_2$ through the relations:
\beq\label{coefficienti}
\nn
c_0=1-\frac{2M}{R_2}\,,
\; c_1=\frac{6M}{R_2}-2\,,
\; c_2=1-\frac{6M}{R_2}\,,
\; c_3=\frac{2M}{R_2}\,,\\
\nn
d_0=-2i\omega R_2+\frac{6M}{R_2}-2\,,
\; d_1=2\left(1-\frac{6M}{R_2}\right)\,,
\; d_2=\frac{6M}{R_2}\,,\\
\nn
e_0=\frac{6M}{R_2}-l(l+1)\,,
\; e_1=-\frac{6M}{R_2}\,.
\eeq
Let us now perform a power-series expansion of $\phi(z)$:
\be\label{expans}
\phi(z)=\sum_{n=0}^\infty a_n z^n\,.
\ee
By substituting this expression in Eq.~(\ref{stellLeav}), the expansion
coefficients $ a_n$ are found to satisfy a four-term recurrence relation of
the form:
\beq
\label{4trrgiusta}
&&\alpha_1 a_{2}+\beta_1 a_1+\gamma_1 a_{0}=0\,, \quad n=1\,,\\
&&\alpha_n a_{n+1}+\beta_n a_n+\gamma_n a_{n-1}+\delta_n a_{n-2}=0\,,
\quad n\geq 2\,,
\nonumber
\eeq
where:
\beq\label{primostep1}
&&\alpha_n=n(n+1)c_0\,,\quad n\geq 1\,,\\
\nn
&&\beta_n=(n-1)n c_1+n d_0\,,\quad n\geq 1\,,\\
\nn
&&\gamma_n=(n-2)(n-1)c_2+(n-1)d_1+e_0\,,\quad n\geq 1\,,\\
\nn
&&\delta_n=(n-3)(n-2)c_3+(n-2)d_2+e_1\,,\quad n\geq 2\,.
\eeq
The coefficient $a_0$ is a normalization constant, and it is irrelevant from
the point of view of imposing outgoing-wave boundary conditions. The ratio
$a_1/a_0$ can simply be determined by imposing the continuity of $ \Psi$ and
$\Psi'$ at $r=R_2$, since from Eq.~(\ref{subst}) it follows that:
\beq
\label{a0fc}
&&a_0=\left.\phi\right|_{z=0}=\frac{\Psi(R_2)}{\chi(R_2)}\,,\\
&&\frac{a_1}{a_0}=\frac{R_2}{\Psi(R_2)}
\left[\Psi'(R_2)+\frac{i\omega R_2}{R_2-2M}\Psi(R_2)\right]\,.
\label{a1a0}
\eeq
In the axial case, the values of $\Psi(R_2)$ and $\Psi'(R_2)$ can be obtained
by the taking the interior solution (\ref{eq:interior_sol}) at $r=a_-$
and applying the junction conditions (\ref{junctionaxial}) to determine the
wavefunction in the exterior, i.e., at $r=a_+$. From then onwards, we
can numerically integrate the Regge-Wheeler equation (\ref{masterSch}) up to
$r=R_2$.  The remaining coefficients can then be determined by recursion from
Eq.~(\ref{4trrgiusta}).
In the polar case we proceed in a similar way: we obtain the Zerilli function
$Z^{\rm out}$ and its derivative at $r=a_+$ by imposing the matching
conditions (\ref{junctionpolar1})--(\ref{junctionpolar3}). Then we use
Eq.~(\ref{chandratr}) to obtain the corresponding Regge-Wheeler function at
$r=a_+$, integrate forwards to find $a_0$ and $a_1$, and finally
obtain the remaining coefficients by recursion.

To apply the continued fraction technique, it is easier to consider
three-term recurrence relations.  Leaver has shown that the four-term
recurrence relation (\ref{4trrgiusta}) can be reduced to a three-term
recurrence relation by a gaussian elimination step \cite{Leaver:1990zz}.  
Define:
\be\label{posizioni}
\hat{\beta}_0=\frac{a_1}{a_0}\,, \qquad
\hat{\alpha}_0=-1\,,
\ee
where $a_1/a_0$ is obtained numerically from Eq.~(\ref{a1a0}). Now set:
\beq\label{gausselim1}
&&\hat{\alpha}_n=\alpha_n\,,\quad
\hat{\beta}_n=\beta_n\,,\quad (n=0,1)\,,\nonumber\\
&&\hat{\gamma}_n=\gamma_n\,,\quad (n=1)\,,
\eeq
and for $n\geq 2$:
\beq\label{gausselim2}
&&\hat{\alpha}_n=\alpha_n\,,
\qquad
\hat{\beta}_n=\beta_n-\frac{\hat{\alpha}_{n-1}\delta_n}{\hat{\gamma}_{n-1}}\,,
\nn\\
&&\hat{\gamma}_n=\gamma_n-\frac{\hat{\beta}_{n-1}\delta_n}{\hat{\gamma}_{n-1}}\,,
\qquad \hat{\delta}_n=0\,.
\eeq
By this gaussian elimination, Eq.~(\ref{4trrgiusta}) reduces to:
\be
\hat{\alpha}_n a_{n+1}+\hat{\beta}_n a_n+\hat{\gamma}_n a_{n-1}=0\,.
\ee
The elimination step is not as trivial as it may seem, because in the process
one of the {\it three} independent solutions to Eq.~(\ref{4trrgiusta}) is
lost.  It can be shown that this solution is not relevant for our purposes
\cite{Leins:1993zz}.

We now turn to investigating the asymptotic behavior of the coefficients
$a_n$ in the expansion (\ref{expans}).  Let us make the ansatz:
\be\label{alphan}
\lim_{n\to \infty}\frac{a_{n+1}}{a_n}=1+\frac{h}{n^{1/2}}+\frac{k}{n}+\dots
\ee
Dividing Eq.~(\ref{4trrgiusta}) by $ n^2 a_n$, keeping terms up to $\sim
n^{-3/2}$ and equating to zero the various terms in the expansion in powers of
$ n^{-1/2}$ we find the relations:
\beq
&&c_0+c_1+c_2+c_3=0\,, \\
\nn
&&2c_0+c_1-c_3=0\,, \\
\nn
&&h^2=2i\omega R_2\,, \\
\nn
&&k=-\frac{3}{4}+i\omega (R_2+2M)\,.
\eeq
The first two of these equations are identities.  Substituting the second pair
of equations in Eq.~(\ref{alphan}) we get:
\be\label{angrande}
\lim_{n\to \infty} a_n=n^{-3/4+i\omega (R_2+2M)}e^{\pm 2\sqrt{2i\omega R_2 n}}\,.
\ee
According to a definition given by Gautschi \cite{Gautschi}, the solution of
Eq.~(\ref{alphan}) corresponding to the plus sign in Eq.~(\ref{angrande}) is
said to be {\it dominant}, whereas that corresponding to the minus sign is
said to be {\it minimal} \cite{Gautschi}. If we select the minimal solution
the expansion (\ref{expans}) is absolutely and uniformly convergent outside
the star, provided that we choose $R_2$ such that $R_2/2<a<R_2$ and
$R_2>2$. Furthermore, according to Eq.~(\ref{subst}), the solution to
Eq.~(\ref{mastereq}) behaves as a pure outgoing wave at infinity, i.e. it is
the a QNM wavefunction.  Thus, the key point is to identify the minimal
solutions of Eq.~(\ref{alphan}).  According to a theorem due to Pincherle
\cite{Gautschi}, if Eq.~(\ref{alphan}) has a minimal solution then the
following continued fraction relation holds:
\be\label{LNSmatch}
\frac{a_1}{a_0}=\frac{-\hat{\gamma}_1}{\hat{\beta}_1-}
\frac{\hat{\alpha}_1\hat{\gamma}_2}{\hat{\beta}_2-}
\frac{\hat{\alpha}_2\hat{\gamma}_3}{\hat{\beta}_3-}
\dots
\ee 
where the continued fraction on the RHS is convergent and completely
determined since the coefficients $\hat{\alpha}_n$, $\hat{\beta}_n$ and
$\hat{\gamma}_n,$ defined in eqs. (\ref{gausselim1}),(\ref{gausselim2}) are
known functions of $\omega$.  Moreover, from Eqs.~(\ref{a1a0}) and
(\ref{posizioni}) it is apparent that the dependence on the stellar model is
all contained in the ratio $a_1/a_0$.
Keeping in mind the definitions (\ref{posizioni}), Eq.~(\ref{LNSmatch}) can be
recast in the form:
\be\label{cuore}
0=f_0(\omega)=\hat{\beta}_0-
\frac{\hat{\alpha}_0\hat{\gamma}_1}
{\hat{\beta}_1-}\frac{\hat{\alpha}_1\hat{\gamma}_2}{\hat{\beta}_2-}
\frac{\hat{\alpha}_2\hat{\gamma}_3}{\hat{\beta}_3-}\dots\\
\ee
Using the inversion properties of continued fractions \cite{Wall}, the latter
equation can be inverted $ n $ times to yield:
\beq\label{cuore2}
0=f_n(\omega)&=&\hat{\beta}_n-\frac{\hat{\alpha}_{n-1}
\hat{\gamma}_n}{\hat{\beta}_{n-1}-}\frac{\hat{\alpha}_{n-2}
\hat{\gamma}_{n-1}}{\hat{\beta}_{n-2}-}...\frac{\hat{\alpha}_0
\hat{\gamma}_1}{\hat{\beta}_0} \nonumber\\
& -&
\frac{\hat{\alpha}_n\hat{\gamma}_{n+1}}{\hat{\beta}_{n+1}-}
\frac{\hat{\alpha}_{n+1}\hat{\gamma}_{n+2}}{\hat{\beta}_{n+2}-}
\frac{\hat{\alpha}_{n+2}\hat{\gamma}_{n+3}}{\hat{\beta}_{n+3}-}... \quad
\eeq
for $n=1,2,...$. These $n$ conditions are analytically equivalent to
Eq.~(\ref{cuore}).  However, since the functions $f_n(\omega)$ have different
convergence properties, each of them is best suited to find the quasinormal
frequencies in a given region of the complex $\omega$ plane.  This is the main
reason for the accuracy and flexibility of the continued fraction technique.

%%%%%%%%%%%%%%%%%%%%%%%%%%%%%%%%%%%%%%%%%%%%%%%%%%%%%%%%%%%%%%%%%%%%%%%%%%%%%%%%
\section{\label{app:highC}High-compactness limit}
%%%%%%%%%%%%%%%%%%%%%%%%%%%%%%%%%%%%%%%%%%%%%%%%%%%%%%%%%%%%%%%%%%%%%%%%%%%%%%%%

To investigate the behavior at the surface of the gravastar in the
high-compactness limit, we use the $z \to 1-z$ transformation law for the
hypergeometric function \cite{Abramowitz:1970as},
\begin{eqnarray}
  &&F(a,b,c,z)\nonumber \\ 
  &=& (1\!-\!z)^{c-a-b}
\frac{\Gamma(c)\Gamma(a+b-c)}{\Gamma(a)\Gamma(b)}
 \nonumber \\ 
 &&\times F(c\!-\!a,c\!-\!b,c\!-\!a\!-\!b\!+\!1,1\!-\!z)  \nonumber \\
&  +&
\frac{\Gamma(c)\Gamma(c-a-b)}{\Gamma(c-a)\Gamma(c-b)}
 \,F(a,b,-c\!+\!a\!+\!b\!+\!1,1\!-\!z)\,. \quad
 \label{transformation law}
\end{eqnarray}
Using Eq.~(\ref{eq:interior_sol}) in the limit when $C\to 1$ and $r\to a$ we
get
\beq
\Psi&\approx& \left[\frac{2(a-r)}{a}\right]^{iM\omega}
\frac{\Gamma(l+\frac{3}{2})\Gamma(-i2M\omega)}
{\Gamma(\frac{2+l-i2M\omega}{2})\Gamma(\frac{1+l-i2M\omega}{2})} 
\nonumber \\
&+&\left[\frac{2(a-r)}{a}\right]^{-iM\omega}
\frac{\Gamma(l+\frac{3}{2})\Gamma(i2M\omega)}
{\Gamma(\frac{1+l+i2M\omega}{2})\Gamma(\frac{2+l+i2M\omega}{2})}\,. \quad\quad
\eeq
Within our conventions the first term is in-going, while the second term is
out-going near the surface. So it is clear that in this regime both in- and
out-going modes are present, and QNMs do {\it not} reduce to the Schwarzschild
QNMs (which require only in-going waves).  Furthermore, we can clearly see
that in-going and out-going waves always have the same magnitude: the
gravastar appears like a reflecting object, as suggested
by~\cite{Pfister:1991ky}.  Because this reflection happens in a polar
coordinate system, it can simply be interpreted as due to the fact that waves
going into a lossless gravastar will re-emerge without loss.  Nevertheless,
such a behavior already supports the conclusions of
Ref.~\cite{Cardoso:2007az,Cardoso:2008kj}, which showed that (for scalar
fields) the ergoregion instability is more effective when the surface of the
compact objects behaves like a ``perfect mirror'' in this sense.

It is easy to show that in the high-compactness limit
\be
\Psi'(a_-)=\frac{i\omega\,a}{a-2M}\Psi(a_-)\,.
\ee
Solving for the metric quantities we find, up to dominant terms in $a-2M$,
\beq
K(a_-)&=&\frac{l(l+1)+2ia\omega}{2a}\Psi(a_-)\,, \\
K'(a_-)&=&-\omega\frac{-il(l+1)+2\omega a}{2(a-2M)}\Psi(a_-)\,,\\
H_1(a_-)&=&-\frac{\omega\,a(\omega\,a-i)}{a-2M}\Psi(a_-)\,,
\\
H_1'(a_-)&=&-\frac{\omega(\omega\,a-i)(4M+i\omega a^2)}{(a-2M)^2}
\Psi(a_-)\,,\\
H_0(a_-)&=&-\frac{\omega\,a(\omega\,a-i)}{a-2M}\Psi(a_-)\,,
\\
H_0'(a_-)&=&-\frac{\omega(\omega\,a-2i)(2M+i\omega a^2)}{(a-2M)^2}
\Psi(a_-)\,.
\eeq
In the exterior we get
\beq
K(a_+)&=&\frac{l(l+1)+2ia\omega}{2a}\Psi(a_-)\,,
\\
K'(a_+)&=&-\omega\frac{-il(l+1)+4\omega a}{2(a-2M)}\Psi(a_-)\,,\\
H_1(a_+)&=&-\frac{\omega\,M(i+4M\omega)}{a-2M}\Psi(a_-)\,,
\\
H_1'(a_+)&=&-\frac{M\omega(2M\omega+i)(1-4iM\omega)}{(a-2M)^2}
\Psi(a_-),\qquad\\
H_0(a_+)&=&-\frac{M\omega(i+4M\omega)}{a-2M}\Psi(a_-)\,,\\
H_0'(a_+)&=&H_1'(a_+)\,.
\eeq
Notice that, even though $H_1$ is not continuous at $r=a$, the Zerilli
function is. Indeed, we get
\be
Z^{\rm out}(a_+)=\Psi(a_-)\,,\qquad \Psi(a_-)=\Psi(a_+)\,.
\ee
Thus, we conclude that in the high-compactness limit, the master wavefunction
for polar perturbations is continuous across the shell.  A trivial extension
of the known Schwarzschild results then shows that {\it polar and axial
  perturbations are isospectral} for large compactness, i.e., when $a\to 2M$
and $\mu\to 1/2$.

\end{document}